\documentclass[12pt,preprint]{aastex}
\shorttitle{Proper Motions in the Galactic bulge}
\shortauthors{Vieira, et al.}

\begin{document}

\title{Proper Motions in the Galactic Bulge: Plaut's Window}

\author{Katherine Vieira, Dana I. Casetti-Dinescu,}
\affil{Astronomy Department, Yale University, P.O. Box 208101, New Haven, CT 06520}
\email{vieira,dana@astro.yale.edu}

\author{Ren\'e A. M\'endez,}
\affil{Depto. de Astronomia, Universidad de Chile, Casilla 36-D, Santiago, Chile}
\email{rmendez@das.uchile.cl}

\author{R. Michael Rich}
\affil{Division of Astronomy, University of California, 8979 Math Sciences, Los Angeles, CA. 90095-1562}
\email{rmr@astro.ucla.edu}

\author{Terrence M. Girard, Vladimir I. Korchagin and William van Altena}
\affil{Astronomy Department, Yale University, P.O. Box 208101, New Haven, CT 06520}
\email{girard,vik,vanalten@astro.yale.edu}

\author{Steven R. Majewski}
\affil{Department of Astronomy, University of Virginia, P.O. Box 400325, Charlottesville, VA 22904-4325}
\email{srm4n@didjeridu.astro.virginia.edu}

\author{Sidney van den Bergh}
\affil{National Research Council of Canada, Herzberg Institute of Astrophysics,
5071 West Saanich Road, Victoria, BC V9E 2E7}
\email{sidney.vandenBergh@nrc-cnrc.gc.ca}

\begin{abstract}

A proper motion study of a field of $20'\times 20'$ 
inside Plaut's low extinction window (l,b)=($0^o,-8^o$), has been completed.
Relative proper motions and photographic $BV$ photometry 
have been derived for $\sim 21,000$ stars reaching to $V\sim 20.5$ mag,
based on the astrometric reduction of 43 photographic plates, 
spanning over 21 years of epoch difference.
Proper motion errors are typically 1 mas yr$^{-1}$ and 
field dependent systematics are below 0.2 mas yr$^{-1}$.

Cross-referencing with the  2MASS catalog yielded a sample of $\sim 8700$
stars, from which predominantly disk and bulge subsamples were selected photometrically
from the $JH$ color-magnitude diagram. The two samples exhibited different
proper-motion distributions, with the disk displaying the expected
reflex solar motion as a function of magnitude. Galactic rotation
was also detected for stars between $\sim$2 and $\sim$3 kpc from us.
The bulge sample, represented by red giants, has an intrinsic proper motion dispersion of
$(\sigma_l,\sigma_b)=(3.39, 2.91)\pm (0.11,0.09)$ mas yr$^{-1}$,
which is in good agreement with previous results, and indicates
a velocity anisotropy consistent with either rotational broadening or tri-axiality.
A mean distance of $6.37^{+0.87}_{-0.77}$ kpc has been estimated for
the bulge sample, based on the observed $K$ magnitude of the horizontal
branch red clump. The metallicity $[M/H]$ distribution was also obtained
for a subsample of 60 bulge giants stars, based on calibrated photometric indices. 
The observed $[M/H]$ shows a peak value
at $[M/H]\sim -0.1$ with an extended metal poor tail and around 30\% of the
stars with supersolar metallicity. 
No change in proper motion dispersion was observed as a function of $[M/H]$.
We are currently in the process of obtaining  CCD $UBVRI$ photometry for the
entire proper-motion sample of $\sim 21,000$ stars.

\end{abstract}


\section{Introduction}

The bulge of the Milky Way remained hidden for a long time, 
due to obscuration by dust, until radio, infrared, X-ray and gamma-ray observations 
revealed some of its properties.
Despite substantial efforts to probe its structure, dynamics
and stellar populations, a clear evolutionary picture
for the bulge is still not in hand. 
The stellar populations of the triaxial
bulge form a dynamically old and relaxed structure 
in equilibrium, however, the bulge might still accrete debris from
an infalling dwarf galaxy, and can also be affected by the influence
of the bar, which was presumably formed by a disk instability.
Last but not least, there is observational evidence for recent star
formation at the very nucleus of our Galaxy.
Hence, studies of the age and metallicity distribution of the bulge stars
are fundamental to understanding the structure and formation of all galaxies,
including the Milky Way. Our galaxy's bulge plays a critical role in this
regard, because it is the nearest central spheroid close enough to allow
detailed studies of the individual stars. This makes possible as well
to perform a close inspection of the kinematics in the center of our
galaxy, that can be used to model the gravitational potential and
the distribution of mass, visible and dark.

The bulge can be described as an oblate
ellipsoid, with a 2 kpc major axis and axis ratios of 1.0:0.6:0.4
(Rich 1998). It consists of mostly old stars (ages $>$ 9 Gyr) with a range of metallicities 
($-1<[Fe/H]<0.5$) and average metallicity a little under solar ($[Fe/H]=-0.2$). 
Recent studies show that the bulge has higher [O/Fe] than both thick and thin disk
(Zoccali et al. 2006), which suggest that it formed as a prototypical old spheroid,
before the disk and more rapidly.
Within the bulge there are about $2\times 10^{10}M_\odot$ of stars that produce a luminosity 
of $L\sim 5\times 10^9 L_\odot$ (Sparke \& Gallagher 2000, pp. 25),
and a supermassive black hole with a mass of $4\times 10^6 M_\odot$ (Ghez et al. 2005). 
Recent star formation has been observed in the inner part of the bulge,
possibly induced by the bar funneling gas into the nuclear region.
Coexisting with the bulge is the bar, a long thin structure of 4 kpc
half length and $\sim 100$ pc scale height, constrained to the plane at 
a position angle of $43^o.0\pm 1^o.8$, as revealed by the red clump stars observed
along different lines of sight in the inner Galaxy.
In contrast, the bulge is a fatter triaxial structure with a $\sim 500$ pc
scale height and a position angle of $12^o.6\pm 3^o.2$ (Cabrera-Lavers et al. 2007).
 
Low extinction windows towards the Galactic bulge provide the only 
possible opportunity to measure the kinematics of the stars in the optical range,
either through radial velocities or proper motions. 
Baade's Window at (l,b)=(1$^o$,-4$^o$) is the best known 
example of such a field, and a list of several others can be found in
Dominici et al. (1999), Blanco \& Terndrup (1989) and Blanco (1988). 
These low-extinction windows are still limited by dust extinction and 
large zenith distance as seen from northern hemisphere, so that only a few studies 
involving radial velocities and proper motions have been made.

For the radial velocity investigations, several measurements
have been made. For example, Tyson \& Rich (1991) used optical spectra from 33 carbon stars, 
and found a radial velocity dispersion $\sigma_r=110 \pm$ 14 km s$^{-1}$,
which decreased from 131 km s$^{-1}$ at low Galactic latitude to 81 km s$^{-1}$ 
for the higher latitudes. Minniti (1996) measured $\sigma_r=71.9\pm$ 3.6 
km s$^{-1}$ using 194 bulge K giants located at $(l,b)=(8^o,7^o)$.
More recently, Rich et al. (2007) presented the first results of an
ongoing large scale radial velocity survey of the Galactic bulge
using M giants. They observe a gradient in $\sigma_r$ vs. Galactic longitude,
that goes from about 75 km s$^{-1}$ at $|l|=10$ up to about 110 km s$^{-1}$
at $l=0$. Other works, like Kohoutek \& Pauls (1995) and Zijlstra et al. (1997),
relied on planetary nebulae (PNe) as the most luminous objects among the old stellar
population to trace the kinematics of the bulge.
Zijlstra et al. (1997) found $\sigma_r=114$ km s$^{-1}$, 
though they say that a bias may exist in their sample
since many of their PNe are located far from the bulge.
Beaulieu et al. (2000) has analyzed a sample of 373 PNe and measured
radial velocities dispersion, which ranged from about 60 km s$^{-1}$ to
110 km s$^{-1}$ for $|l|<15$ , but were smaller for
fields beyond that region in galactic langitude.
It is still unclear if these observed trends reflect 
a gradient in the bulge rotation or are just produced by disk contamination.

The first proper motion investigation of the 
Galactic bulge was made by Spaenhauer et al. (1992)
and since then a few additional studies have been made, as we will detail later
in this paper. All proper motion investigations of the bulge,
including the one we present in this paper, more or less agree on a velocity
dispersion based on proper motions of around 
$\sigma_l=$150 km s$^{-1}$ and $\sigma_b=$110 km s$^{-1}$ in Galactic longitude
and latitude respectively. The observed anisotropy $\sigma_l/\sigma_b$
has been explained by the effect of bulge rotation on the observed stellar velocities,
which produces an apparent increase of $\sigma_l$ in an
intrinsically isotropic oblate rotator body (Zhao et al. 1996).
It is also found that more metal rich stars have a larger
apparent anisotropy $\sigma_l/\sigma_b$, a smaller $\sigma_b$, and 
$\sigma_r>\sigma_b$ when compared to metal poor stars.
This latter result has been explained as 
evidence of a fast-rotating and intrinsically anisotropical bar component
(Soto et al. 2006). As suggested by Zhao et al. (1996), the bulge probably consists of several
co-spatial stellar populations with different kinematics and/or density
distributions, but on average the whole sample turns out to agree well with
the oblate rotator model.

One of the least studied low extinction windows is the so-called
Plaut's window, at (l,b)=(0$^o$,-8$^o$)\footnote{
$(RA,DEC)=(18^h18^m8\stackrel{^s}{_.}32,-32^o51' 35\stackrel{"}{_.}7)$},
named after Lukas Plaut, who was involved 
for many years in the search and study of RR Lyrae variables
located in relatively unobscured regions in the Sagittarius-Scorpius section,
when working in the Groningen-Palomar Variable Star Survey (Plaut 1973, Blaauw 2004).
One of the many results of his work, was finding that the distance 
to the Galactic center was around 8.7 kpc,
less than the value of 10 kpc generally assumed at that time (Oort \& Plaut 1975).

Plaut's low extinction window
($E(B-V)=0.25\pm 0.05$ mag, van den Berg \& Herbst 1974),
has smaller reddening and is less crowded than Baade's window,
giving us the opportunity to study the kinematical
properties of the bulge minor axis, at approximately 1 kpc south of the Galactic center,
in the transition zone between the metal-rich Galactic bulge and
the metal-poor halo.  In a preliminary study, M\'endez et al. (1996) derived 
relative proper motions from 5 of our 43 plates,
and measured the proper motion dispersion for all stars with $V\leq 18$,
without making any particular selection for bulge stars.
Their result sought to demonstrate the potential of the plate material
for obtaining accurate proper motions, if all 43 plates
available were used. In this paper we utilize all
43 plates to create a catalog of relative proper motions and
photographic $BV$ photometry, for $\sim$ 21,000 stars in a 20'$\times$20'
field of view.

\section{The Plate Material}

This investigation is based on 43 photographic plates, taken with 
five different telescopes: 
the 51-cm Double Astrograph at the Yale Southern Observatory (YSO) in
El Leoncito, Argentina, the 2.1-m at KPNO, the du Pont 2.5-m  at Las Campanas, 
the Blanco 4-m at CTIO and the Hale 5-m  at Palomar Observatory.
The plates were taken in both $B$ and $V$ passbands with a seeing ranging from
1-3 arcsec and their characteristics are listed in Table \ref{plates}. 
Some plates are deep enough ($V_{lim}\sim 22$) to reach below the main-sequence turnoff 
of the bulge.
The size and layout of the different plates can be seen in Figure 
{\ref{fov_borders}.

The plates were digitized with the Yale
PDS 2020G microdensitometer in raster-scan mode (i.e., the entire plate
was scanned rather than scanning an input list of objects), 
under the best possible thermal conditions
to avoid instrumental drifts during the 8-10 hour scans. 
The aperture size (and therefore pixel size)
was defined so that the scale in mas per pixel would
be similar for all the plates. 
The KPNO and Hale plates serve as the first-epoch material, the du Pont plates
from 1979 we call intermediate-epoch plates, and the rest of the
photographic material provide second-epoch plates.

\section{The Astrometric Reduction}

\subsection{Object Detection and Centering Precision}

Objects were detected in the scans of each plate using the SExtractor code (Bertin \& 
Arnouts 1996). The preliminary centroids of objects computed with 
the SExtractor code were used as input positions into the Yale 2D Gaussian
centering routine (Lee \& van Altena 1983)
thus refining the centering precision.
The positions thus obtained on each plate were transformed into 
the Digital Sky Survey with the purpose of obtaining 
lists of objects on the same equatorial system. These lists 
were then cross-correlated to obtain a master list in which each object is
assigned a unique identification number. The matching radius was chosen
to be 2 arcseconds.
The centering precision per star and single measurement for well-measured stars
is listed in Table \ref{plates}. This number was estimated from the
unit weight error of coordinate transformations of pairs of plates
taken with the same telescope at the same epoch.

\subsection{Photographic Photometry}

We have obtained $BV$ photographic photometry by calibrating 
the instrumental magnitudes from each plate with 
26 photoelectric standards from van den Bergh \& Herbst (1974).
The magnitude of each object is given by the mean of 
all determinations and its formal error is computed from the scatter of
the determinations. Outliers are eliminated in an iterative process
that is based on the estimated scatter of the measurements.
The $BV$ color-magnitude diagran (CMD) and the photometric 
errors in the visual passband as a function of $V$ magnitude
are shown in Figure \ref{cmd_err}, left and right panels respectively.
The solid line in the right panel of Fig. \ref{cmd_err} represents
the moving median of the $V$-magnitude errors. 
Thus, the formal photometric errors are $\sim 0.1$ mag for well-measured stars
and between 0.2 and 0.3 mag for stars fainter than $ V = 18 $.

Due to the high reddening, photometry in the bulge is typically
done in the $V$ and $I$ bands. The only CCD $BV$ photometry available in the bulge 
is given by Terndrup (1988) in both the Baade and Plaut windows.
By comparing the CMD from Terndrup (1988, their Fig. 4) with ours,
we find qualitative agreement for the main features:
the giant branch of the bulge ($B-V>0.85$ and $V<17$), 
the main sequence of foreground disk stars ($B-V<0.85$ and $V<17$),
and a low-density, blue horizontal branch of metal poor stars in the 
bulge ($V\sim 16$ and $-0.1\leq B-V \leq 0.4$).
   
Our main reason for obtaining calibrated photographic photometry 
is to have  $(B-V)$ colors that are crucial in modeling color
terms in the astrometric reductions. Color terms are generally due to
differential atmospheric color refraction which, in this work, is very large
for the 1st epoch plates, because they were taken in the northern hemisphere
at zenith distances of $\sim 60\arcdeg$.

\subsection{Determination of Relative Proper Motions}

The proper motions were computed using the central plate overlap method
described in Girard et al. (1989). This procedure
provides differential proper motions by modeling all plates into
one master plate in an iterative process. The stars that are used to
compute the plate model span a given magnitude and color range, and 
they define the reference system. For our case, these stars are a collection of
disk and bulge stars; therefore the proper motions are on a 
relative reference system. Only a few galaxies were identified on
a handful of plates, and thus, at this point, we are not able to 
provide a good measurement for the zero point of the proper-motion system
(i.e., provide absolute proper motions).

Before modeling the plates into the reference plate, a set of 
corrections was applied to the positions.
First, we corrected the positions for optical field angle distortion (OFAD)
for the plates taken with telescopes with previous astrometric studies.
Thus the CTIO, duPont and Hale plates
were precorrected using the distortion coefficients from 
Chiu (1996) and from Cudworth \& Rees (1991). The center of distortion
was taken to be the tangent point of each plate.

The second correction is the well-kown magnitude equation, a 
systematic bias introduced during long exposures due to inaccurate guiding and
the nonlinearity of the photographic plates (see e.g., Kozhurina-Platais et
al. 1995; Guo et al. 1993; Girard et al. 1998). In this study
there are neither stars with negligible intrinsic proper-motion dispersion
such as stars in clusters which are used to model the proper-motion
magnitude equation (e.g., Guo et al. 1993),
nor multiple images of different magnitudes 
per single star, such as the Southern Proper-Motion Program 
(Girard et al. 1998) uses to correct its magnitude
equation. Therefore, our approach to model and correct for
magnitude equation is different than in the above-mentioned studies, and
has to rely on an assumption about the collection of plates.
The assumption is that for a large set of plates taken with the
same telescope at the same epoch, the variation in magnitude equation 
is random and therefore the average magnitude equation over the whole
set of plates is zero. In our case, this approach is suitable for the
modern du Pont plates, and for the old KPNO plates as these sets 
have many plates (Table \ref{plates}) that can help build statistics.
Thus, we have first considered the 
modern du Pont plates. All modern du Pont plates were transformed into
one from the same set (what we call the `assigned plate'),
which is the deepest, and taken at the
smallest hour angle. The plate model is a polynomial that has up to
4th order terms in the coordinates, and a linear color term when needed. A plot of 
residuals as a function of magnitude reflects the difference
in magnitude equation, between the plate being studied and the assigned plate.
The residuals as a function of magnitude were 
determined for all modern du Pont plates (except for the assigned plate itself) 
and then averaged to determine a
single curve which, under our assumption, represents the assigned plate's 
magnitude equation. Once the assigned plate was corrected for it,
all other du Pont plates were placed onto this presumably magnitude equation-free system. 
A similar procedure was applied 
to the KPNO plates. We then calculated a preliminary catalog 
of proper motions based only on the modern du Pont plates and the old
KPNO plates. This proper-motion system is presumably free of 
magnitude equation.
Subsequent iterations of the proper motion catalog, 
included the intermediate-epoch du Pont 
plates, then the  CTIO and YSO plates, and finally the Hale plates.
Magnitude trends for these latter sets of plates  determined from their
mapping into the preliminary magnitude-equation free system were 
interpreted as magnitude equation and 
corrected for accordingly. The correction was determined and applied 
for each individual plate, before they were introduced into the 
proper-motion solution. 
 
The proper motion for each star was computed using a least-squares 
linear fit of position as a function of time.  Measures 
that differed by more than 0.2 arseconds from the best-fit
line were discarded from the solution. Individual proper-motion errors
were computed from the formal error of the slope as determined by the 
scatter of the residuals about the best-fit line.
 
Since the $x,y$ coordinate system is aligned with right ascension and 
declination, the proper motions thus obtained are in the equatorial system. 
These were transformed to the Galactic coordinate system to 
facilitate their kinematical interpretation discussed in the
following section.
Initially our catalog had data for $\sim$ 31,000 stars,
covering $30\arcmin\times30\arcmin$ on the sky, but inadequately modeled
OFAD obvious at the edges of the field restricted the final catalog to 21,660 stars in
an area of $20\arcmin\times20\arcmin$.
As expected, proper-motion errors increase with $V$-magnitude, 
due to lower S/N. In Figure \ref{emum} we show our 
proper-motion errors as a function of magnitude. The solid line 
indicates a moving median for the whole sample. Well-measured
stars have proper-motion errors of $\sim 0.5$ mas yr$^{-1}$.
Internal estimates of the proper motion field dependent
systematics are below 0.2 mas yr$^{-1}$. 

Since our goal is to accurately determine the intrinsic proper-motion
dispersion of the bulge (see following section), we must first determine
an accurate estimate of our proper-motion errors. The values shown in 
Fig. \ref{emum} are formal error estimates determined when calculating the 
proper-motions, and they may not necessarily reflect the true errors.
To better estimate the proper-motion errors we divided our plates into two independent
sets and then created two proper-motion catalogs. 
A comparison of the proper-motion differences between these two 
catalogs, indicated that the true errors are 
larger by 6\% in $\mu_l cos b $, and by 10\% in $\mu_b$ than our formal errors.
We have therefore applied these values as corrections to the formal proper-motion errors,
and these adjusted errors are the ones we list in our catalog.

\section{Results}

A cross-reference was made between our catalog and the 2MASS Catalog (Cutri et al. 2003), 
so we could use the infrared 
photometry to cleanly separate bulge from foreground disk stars.
Since 2MASS does not go as deep as our catalog,
we ended up with 8721 matched stars, which represent $\sim 90\%$ of
the 2MASS detections in the field. For this sample, our (adjusted) proper-motion errors
are typically 0.5 mas yr$^{-1}$ in each coordinate.

Figure \ref{cmd_jh} shows the infrared $J$ vs. $J-H$ diagram for our Plaut's window sample. 
It is similar to that of Zoccali et al. (2003),
for (l,b)=(0.277,-6.167), except for a slight correction
due to differential reddening and extinction. Both CMDs use 2MASS photometry.
In Figure \ref{cmd_jh}, the Red Giant Branch (RGB) is mostly populated by bulge red giants
and can be easily distinguished. There is also visible
what looks like a slightly bluer and brighter thin sequence of a few 
stars running parallel to the RGB stars. Some of these stars
probably belong to the metal-poor population of the bulge, but disk
giants are also present in this part of the CMD.
As pointed out by Zoccali et al. (2003) and seen in our CMD as well, 
the Horizontal Branch (HB) red clump of the bulge can be easily seen at $J\sim$13.5 and $J-H\sim$0.6,
merged with the bottom of the RGB. Due to a combination of differential reddening, metallicity dispersion
and depth effect, the HB has a large magnitude spread.
Just below the HB red clump, a second grouping of points,
the RGB bump, is seen at $J\sim$14 and $J-H\sim$0.6.
Foreground main sequence stars belonging to the disk are located in
the almost vertical sequence extending upwards ($J\leq 16$) 
and bluewards ($0.0\leq J-H \leq 0.45$),
widely dispersed along the line of sight (Ng \& Bertelli 1996). 
Another vertical sequence can be 
distinguished at $J-H\sim$0.6 for $J<$13, which corresponds to the disk red clump of 
stars, dispersed in magnitude as a result of their large spread in distance 
and reddening (Zoccali et al. 2003)

From Fig. \ref{cmd_jh}, we selected 482 bulge stars, by using 
the decontaminated bulge sample of Zoccali et al. 2003 (their figure
6c) as a template. We also separated a sample of 
1851 disk main sequence stars by selecting those 
with $0.0\leq J-H\leq 0.45$ and $J\leq 16$.
Each population is expected to have its own kinematical 
characteristics, and this is indeed observed in Figure \ref{pm_histo},
where the bulge sample easily stands out from the disk samples,
in the $\mu_l\cos{b}$ distribution. 
Since the disk sample is populated by blue stars and the bulge sample by red ones,
we considered the possibility of a residual color term producing
a different mean motion between the chosen samples.
This possibility was discarded since an uncorrected color term of this size
on the individual plates would have been so large that it would have
been obvious, but no such effect was seen.

\subsection{The disk sample}

The disk stars selected from the infrared CMD, have a limited color range
around $B-V\sim 0.7$, therefore we can consider the
observed visual magnitude as a rough indicator of distance. A plot of their proper motions components
as a function of visual magnitude (Figure \ref{pm_disk}) reveals a trend consistent with the 
projection of the reflex solar motion in the observed 
proper motions. 
A residual magnitude 
equation was discarded as the cause of this trend, since the bulge sample, which has visual magnitude 
in the same range [12.5,15.5], does not exhibit any trend in its proper motions.

We then proceeded to compare our disk proper motion data with the expected 
projection of the solar motion on the stellar
proper motions. Simple geometry and distance determines this projection,
as stated by eqs. (31) and (32) from Mignard (2000):
\begin{eqnarray}
\mu_l\cos{b} &=& \frac{u_\odot\sin{l}-v_\odot\cos{l}}{4.74d} \label{sun_l}\\
\mu_b        &=& \frac{u_\odot\cos{l}\sin{b}+v_\odot\sin{l}\sin{b}-w_\odot\cos{b}}{4.74 d} \label{sun_b}
\end{eqnarray}
where $\mu_l\cos{b}$ and $\mu_b$ are given in mas yr$^{-1}$, 
$d$ is the distance to the star in kiloparsecs, and
$(u_\odot,v_\odot,w_\odot)$ are the components of the solar motion
with respect to the Local Standard of Rest (LSR), measured
in km s$^{-1}$.
An arbitrary zero-point must be added to our relative 
proper motions, to account for the mean motion of our 
reference frame with respect to the Sun. After this, the proper motions
will only contain the solar motion plus the stellar motions, with
respect to the LSR.

In Figure \ref{solar_fit}, we plot equations (\ref{sun_l}) and
(\ref{sun_b}) for the predicted values of solar motion by
Dehnen \& Binney (1998) (for a zero dispersion population), Mignard (2000) (for K0-K5 stars) 
and Abad et al. (2003) (for $e_\pi/\pi<0.33$),
to be compared with our data. 
The distance in equations (\ref{sun_l}) and (\ref{sun_b}) has been converted to
an apparent magnitude $V$ by assuming that the stars have an absolute magnitude
of $M_V=5.25$. This value of $M_V$ was taken from the observational
HR diagram of Perryman et al. (1997), for main sequence stars having $B-V=0.7$.
A linear correction of 0.1 mags kpc$^{-1}$ was added to the computed 
apparent $V$ magnitude, to account for extinction effects.\footnote{Total
extinction in the Plaut's Window is $A_V=3.1 E(B-V)=0.775\pm 0.155 \sim 0.8$ mags, following
Cardelli et al. (1989).}

For $\mu_l\cos{b}$, both Abad et al. (2003) and Mignard (2000) produce good fits to our data, 
while there is a disagreement with Dehnen \& Binney (1998).
The main difference in the velocity of the reflex solar motion between these three studies,
is in the $v_\odot$-component (Galactic rotation direction), which is approximately
5 km s$^{-1}$, 15 km s$^{-1}$ and 20 km s$^{-1}$ for the three references cited, respectively.
This velocity component defines the solar motion projection
onto $\mu_l\cos{b}$ at $l=0$, as stated by eq. (\ref{sun_l}).
The apparent change in the solar motion depending on the sample of stars
chosen to measure it, is a well known effect which reflects the fact
that stars have kinematics related to their ages. 
For instance, Dehnen \& Binney (1998) find that $v_\odot\approx 25$ km s$^{-1}$ for stars
with $B-V\sim 0.7$, and putting this value into the above equation
improves the fit significantly. 
All three papers more or
less agree on their values of $u_\odot$ and $w_\odot$, and their predictions
match our points in $\mu_b$ closely.  
For stars fainter than $V\sim 16.7$ ($J\sim$ 15.2), the observed trend in $\mu_l\cos{b}$ changes.
At this distance ($\sim$ 2 kpc), the reflex solar motion has decreased
sufficiently so that Galactic rotation begins to be noticed in $\mu_l\cos{b}$.
We will explore this subject in more detail later in the section 5.2.
No change is seen in $\mu_b$, which is consistent with the expectation that
no vertical motion is present in either the disk or the bulge.

\subsection{The bulge sample}

Regarding the bulge, the intrinsic proper motion dispersion can be
calculated from the observed proper motions (Figure \ref{pm_bulge}), 
corrected by the contribution of the estimated errors.
Following Spaenhauer et al. (1992)
\begin{equation}
\frac{1}{n-1}\sum_{i=1}^n (\mu_i-\bar{\mu})^2 =
\sigma^2 + \frac{1}{n}\sum_{i=1}^n\epsilon_\mu^2 
\label{spanhauer}
\end{equation}
where $\sigma$ is the intrinsic dispersion, $\epsilon_\mu$ is
the (adjusted) proper-motion error estimate, and $n$ is the number
of stars in the sample. For our bulge sample of 482 stars, we derive
an intrinsic dispersion of $(\sigma_l,\sigma_b)=(3.39,2.91)\pm (0.11,0.09)$ mas yr$^{-1}$.
The observed anisotropy $\sigma_l/\sigma_b=1.17 \pm 0.05$ has been
explained as a ``rotational broadening'' due to integration over the line
of sight of an intrinsically isotropic
bulge (Zhao et al. 1996). Nevertheless, theoretically speaking, 
the observed anisotropy could also be produced 
by an intrinsically larger azimuthal dispersion of a nonrotating bulge.

Our result compares fairly well with previous studies (Table \ref{pm_in_PW}),
though it is clear that we obtained the one of the highest dispersions of all 
proper motion investigations of the bulge
(Figure \ref{prev_res}, bold purple point).
An increased value of the observed velocity dispersion could be due
to some contamination from halo stars at Plaut's location.
For example, Minniti (1996) measured $\sigma_r=113.5\pm$14.4 km s$^{-1}$
in a sample of 31 ``pure'' halo giants ([Fe/H]$\leq$-1.5) located
at about 1.5 kpc from the Galactic center at $(l,b)=(8^o,7^o)$. 
Assuming $R_\odot=8$ kpc as the mean distance to these halo giants,
and given that the halo is isothermal (Minniti 1996), then $\sigma_r$ can be translated 
into a proper motion dispersion, which will be $\sim$3 mas yr$^{-1}$.
A few halo stars contaminating the bulge sample may increase the real bulge dispersion.
In our case, several runs of the Besancon Model of the Galaxy (Robin et al. 2003, 2004)
at Plaut's Window position, simulating the data we have observed
and the sample we have selected, suggest that this contamination is in fact zero.

The comparison of all proper motion dispersions of the bulge 
as seen in Figure \ref{prev_res}, relies on
a simple but powerful assumption: that all samples are located at the
same mean distance, usually $R_\odot$, which may well not be the case, since
the bulge has a non-neglibigle depth and its orientation also implies
that we might be observing at different distances according to the
position. For example, a velocity dispersion of 100 km s$^{-1}$ corresponds
to a proper motion dispersion of $\sim$ 2.6 mas yr$^{-1}$ for
a distance of 8 kpc, but for stars at 6 kpc from us, the proper
motion dispersion will be measured as $\sim$ 3.5 mas yr$^{-1}$.
Their difference, 0.9 mas yr$^{-1}$, is far larger than the errors
achieved in the most recent papers about the bulge proper motion dispersion,
so such a difference may well be observed.
Therefore, our high value of $(\sigma_l,\sigma_b)$ could then be explained
if our sample of stars happens to be in the closer side of the bulge.


The only distance indicator available to us in Plaut's Window
is the bulge HB red clump, easily visible at $J\sim 13.49, H\sim 12.87$ and
$K\sim 12.67$ (Figure \ref{red_clumps}). As explained by Ferraro et al. (2006),
the absolute $K$ magnitude of the helium-burning red clump stars is mainly influenced
by metallicity and shows little dependence with age. When working
with relatively old metal-rich populations, it is then possible to use this feature
as a distance indicator. Extensive research on this topic has also been 
done using Hipparcos, OGLE and HST data (Paczynski \& Stanek 1998, Stanek \& Garnavich 1998, 
Udalski et al. 1998). Based on the measurements of six metal-rich 
($-0.9<[Fe/H]<-0.3$) clusters, Ferraro et al. (2006) found 
$M_K^{RC}=-1.40\pm 0.2$, where the uncertainty adopted is a conservative
estimate of the age and metallicity effects for the population used to compute it.

Getting a distance modulus is then straightforward, once the value of
the extinction in the $K$-band is given, which though known to be small, 
could significantly change the measured distance, particularly in the highly extincted fields
close to the Galactic center. 
$A_K$ can be estimated from $A_V=0.775\pm 0.155$ 
and Cardelli et al. (1989)
extinction laws: $A_K/A_V=0.114 \pm 0.04$.
From these, $A_K=0.09 \pm 0.05$ magnitudes, and from the measured
observed $K^{RC}=12.67\pm 0.2$, we can obtain the the distance modulus.
Nevertheless we decided to go a bit further,
since the availability of a catalogue by Frogel et al. (1990),
allowed us to directly measure $A_K$.
Frogel at al. (1990) have a list of 90 M giants located in the Plaut's Window,
with derredened $JHK$ in the CTIO/CIT system, H2O, CO and bolometric
magnitudes, and spectral types. Conversion of their $JHK$ into
the 2MASS system was done through Carpenter (2001) color transformations.
A total of 55 stars were crossmatched\footnote{Thanks to a private communication
with one of Frogel et al. (1990) coauthors, Dr. Donald Terndrup,
since positional information has not been published.}
between our data and Frogel at al. (1990).
The resulting measured extinction in the $K$-band is
$A_K=0.05\pm 0.003$ magnitudes, which is still within the error
bars of the value above, though this one is far more precise and comes
directly from observed stars in the field.
In any case, this additional exercise proved to be good to confirm numbers
and also to obtain spectral types 
for the 55 stars subsample, which are going to be of great value
for future developments of this project. This latter value of $A_K=0.05\pm 0.003$
will be adopted for the rest of this work.

Therefore we conclude that the mean distance modulus of our 482 bulge stars is
\[
K^{RC}-M_K^{RC}-A_K=12.67-(-1.40)-0.05 \pm \sqrt{0.2^2+0.2^2+0.003^2}=14.02 \pm 0.28
\]
or equivalently $\sim 6.37^{+0.87}_{-0.77}$ kpc, which means we are looking
$\sim 885$ pc south of the Galactic plane.
If we assume $R_\odot=7.52$ kpc (Nishiyama et al. 2006), a value that
was also obtained by using the HB red clump as a distance indicator,
then our bulge sample is located $\sim 1.21$ kpc closer than the Galactic rotation axis.
Using the larger and most referred value of $R_\odot=8.00$ kpc, 
stresses the fact that our sample is significantly closer than
$R_\odot$, but this shorter distance 
still makes possible for these stars to be part of the bulge.
Indeed, based on the bulge density model of Zhao (1996)\footnote{Zhao (1996) uses
a triaxial structure with a gaussian radial profile and boxy intrinsic shape, 
to describe the bulge. He calls this model {\it bar}, but it is what we currently
call a triaxal bulge. The scale lenghts in Zhao (1996) are 
$a=1.49$, $b=0.58$ and $c=0.40$ kpc in the $x$, $y$ and $z$ axis respectively, assuming
$R_\odot=8$ kpc. In his model, the bulge has a position angle of 20$^o$, ``close to the value
of 13.4$^o$ found by Dwek et al. (1995)'' and it has an
effective axis ratio of 1:0.6:0.4.}, it is straightforward to compute that
the bulge density at the location of our sample is $\sim$ 8\% 
of the highest central value.
Additionally, if we use $d=6.37$ kpc to convert the proper motion dispersion into
a velocity dispersion, we obtain $(\sigma(v_l),\sigma(v_b))=(102.30,87.76)
\pm (3.32,2.85)$ km s$^{-1}$, which compares well with the prediction
by Zhao (1996) of $(\sigma(v_l),\sigma(v_b))\sim(115,80)$ km s$^{-1}$, as
obtained from his Figure 6 (lower panel).

The measured velocity dispersions can also be used to estimate the rotational
velocity at Plaut's location. Using
Zhao et al. (1996) equations (1) and (2), we obtain $52.57\pm 8.02$ km s$^{-1}$.
This value is in agreement with the bulge rotation curve as measured by
several authors, including for example Izumiura et al. (1994), who
used SiO masers at $7^o<|b|<8^o$ and measured
$V_{rot}=50.8\pm 14.2$ km s$^{-1} R$ kpc$^{-1}$, 
where $R$ is the distance from the Galactic rotation axis, and also
Tiede \& Terndrup (1999), who obtained $V_{rot}=53.3\pm 9.3$ km s$^{-1} R$ kpc$^{-1}$,
using stars with $V-I>1.2$ and $I<16.5$, located at $l=-8,+8$ and $b=-6$.
More recent results by Rich et al. (2007) suggest though,
that the bulge may not have a solid
body rotation. Radial velocities of M giants
observed on a strip of fields at $b=-4$, indicate that for $|l|<3^o$, the
bulge rotation curve has a slope of roughly 100 km s$^{-1}$ kpc$^{-1}$,
and beyond that region in galactic longitude, the rotational velocity
flattens out to about $\sim 45$ km s$^{-1}$. Since this last result suggests
a much higher rotation speed at a lower $|b|$, then a possible gradient in
the bulge rotation could explain all these observational results in the
same scenario. In any case, this subject is still not resolved. 
If we consider for example, Izumiura et al. (1995) results using
SiO masers at $|l|<15^o$ and $3^o<|b|<15^o$, excluding strips at $4^o<|b|<5^o$ 
and $7^o<|b|<8^o$, they found a higher rotational rate
of $V_{rot}=73.5$ km s$^{-1} R$ kpc$^{-1}$. Disk contamination and/or small numbers
has been considered as a factor to explain such high rates (Rich et al. 2007)

\subsubsection{Bulge BHB stars}

In addition to the RGB bulge sample,
we also explored the kinematics of another stellar population
that lives in the bulge, the blue horizontal branch (BHB)
of metal poor stars. They can be seen in the optical $BV$ CMD of Figure 
\ref{cmd_err}, as a low-density feature at $-0.5<B-V<0.45$ and $15<V<17$,
comprising a total of 103 stars. Their proper motion distribution looks
a bit noisier but very similar to the observed for the bulge RGB stars sample in Figure 
\ref{pm_histo}, which makes a very good case for these
stars being bona fide members of the bulge, despite the fact that this sample is 
almost five times smaller than the bulge RGB one.  
Three of the BHB stars have very high proper motions. Once these 
stars are discarded, the measured proper motion dispersion is 
$(\sigma_l,\sigma_b)=(4.23,3.48)\pm (0.30,0.25)$ mas yr$^{-1}$,
a result which is about 6 to 7 sigmas away from the dispersion obtained 
with the bulge RGB stars. Halo contamination, more probable in this
part of the CMD, could be responsible for this higher value. 
The observed anisotropy is $\sigma_l/\sigma_b=1.22 \pm 0.12$, which
is within the error bars of the anisotropy measured with bulge RGB stars. 
Taking this number as a face value indicates that proper motion anisotropy
does not change with metallicity, as observed by 
Spaenhauer et al. 1992.  These numbers must be taken with caution though,
since several factors affect them, including small numbers, halo
contamination and even MS disk stars contamination, since
our optical photographic photometry is not precise enough to select a clean sample.

\subsection{Metallicity distribution in the bulge: proper motions vs. [M/H]}

We have also explored the metallicity distribution of our bulge sample, by using
the method explained in Zoccali et al. (2003). A family of hyperbolas in the
plane $(M_K,(V-K)_0)$ represent the upper RGBs of globular clusters that have
a range of metallicities. An inversion of the hyperbolas produces a value
of metallicity for each star. This method required  using our $V$ photographic magnitude
with the $K$ magnitude from 2MASS, and despite the lower precision of
our optical photometry, the dispersion in $V-K$ was similar to the
one observed in Figure 12 of Zoccali et al. (2003), which uses
CCD data alone.

Due to the high sensitivity of $V-K$ to metallicity,
it is very important to make the best possible reddening correction.
Once again, based on Cardelli et al. (1989) extinction law
for $A_V$, and our measurement of $A_K$, we obtain:
$E(V-K)=A_V-A_K=3.1*E(B-V)-0.05=3.1*0.25-0.05=0.73$ magnitudes.
Coming back to Zoccali et al. (2003),
we have applied the corresponding inversion of their equation (1),
to get $[M/H]$ for each star in the bulge sample. Following the recommendations
of that paper, only stars fainter than $M_K=-4.5$ and $(V-K)_0>2.8$ were
considered for the inversion, which left us with a subsample of 60 stars.
Results can be seen in Figure \ref{hyperbolas},
that can be compared with Figures 9 and 12 from Zoccali et al. (2003). 
Despite our small number statistics, the similarity between the 
plots is very good. A histogram of the metallicities obtained is shown
in Figure \ref{metal}, displaying the distribution peak at $[M/H]\sim -0.1$
with a somewhat extended metal poor tail, and $\sim 30\%$ of the stars
with supersolar metallicity, i.e. $[M/H]>0$.
Again, the results are similar to those observed in Zoccali et al. (2003),
which are plotted in gray. Since the slightly bluer and brighter sequence of
few stars parallel to the RGB visible in our Figure \ref{cmd_jh} 
is not included in the bulge sample, it is
possible that some few bulge metal poor stars are missing in Figure \ref{metal}.
If they are included, the overall distribution of [M/H] does
not change significantly, since these few stars - if calibrated using the hyperbolas
method described above - have $-1<[M/H]<-0.3$, and they would smooth
the histogram in that section of it. This would in effect make our histogram
more alike to the one obtained by Zoccali et al. (2003). 

Proper motions as a function of $[M/H]$ can be seen in Figure \ref{mu_mh}.
No significative change in the proper motion dispersion is noticed 
between the metal-poor and the metal-rich stars, but since our sample
is small and our abundances are based on photometric indices,
we do not take this result as a conclusive one in this regard.
The only previous work that explored bulge proper motion
dispersion as a function of metallicity, is Spaenhauer, Jones \& Withford (1992),
who studied a sample of 57 K and M giant bulge stars, separated in two samples:
23 stars with $[Fe/H]\leq 0$ and the other 34 with $[Fe/H]>0$. 
They found some indication that the metal-rich sample had a larger $\sigma_l$ 
and a smaller $\sigma_b$, than the metal-poor sample, though they also mention 
that more data would be ideal to confirm such result. Investigations of the
radial velocity distribution as a function of abundance (Rich 1990, Minniti 1996)
have found clear correlations, that other authors (Tiede \& Terndrup, 1997) impute
to contamination by disk giants. More recently, Soto et al. (2007) explored
this subject by combining the proper motions from Spaenhauer et al. (1992), the
abundances of Terndrup et al. (1995) and the radial velocities of Sadler et al.
(1996), for a total of 315 K and M giants. Their results confirm the trend 
in $\sigma_b$, but see nothing in $\sigma_l$ nor in $\sigma_r$. They found though,  
a significative vertex deviation in the radial velocity vs. $\mu_l$ plane, consistent
with the presence of a bar. On this whole issue, an improved selection of
BHB stars from our catalog, based on CCD photometry, would be ideal to study
their kinematics and better assess if these observed changes are indeed real or not. 
Our paper provides the first kinematic link ever between BHB and metal 
rich bulge stars. 

We certainly agree that extreme caution
must be taken when photometrically selecting bulge stars, because even when kinematic 
information is available to improve the selection of bona fide bulge members,
confusion is still very possible, particularly when observing deep into the Galactic
center. In the next section we will thoroughly discuss this issue.

\section{Discussion}

\subsection{Proper motion dispersion at different locations on the bulge}
The recent availability of proper motion data in numerous locations on the bulge,
and the observation of what looks like trends in proper motion dispersion 
according to $(l,b)$ (Kozlowski et al. 2006, Rattenbury et al 2007),
have prompted the idea that these results may reflect intrinsic 
kinematic features of the bulge. 
For example, Kozlowski et al. (2006) found a systematic variation of $\sigma_l$ vs. $b$
as measured by bulge main sequence (MS) stars
(Figure \ref{prev_res}, upper panel right, blue points). Previous results by 
results by Zocalli et al. (2001) and Kuijken \& Rich (2002), 
both using MS bulge stars as well, do not deviate significantly 
from the trend observed by Kozlowski et al. (2006) 
(Figure \ref{prev_res}, upper panel right, green points).
On the other hand, Spaenhauer et al. (1992) and Feltzing \& Johnson (2002),
who used RGB stars as tracers of the bulge, found larger values of $\sigma_l$
that significantly deviate from the observed trend above, and
are in agreement with our high value of $\sigma_l$, within the error bars
(Figure \ref{prev_res}, upper panel right, yellow points). More recently, Rattenbury et al. 2007,
using a sample -supossedly- dominated by bulge HB red clump giants,
have also found correlations in $\sigma_l$ vs. $b$ 
(Figure \ref{prev_res}, upper panel right, red points), that follow 
Kozlowski et al. (2006) results . On the other hand, Rattenbury 
et al. 2007 also observe a trend in $\sigma_l$ vs. $l$ 
(Figure \ref{prev_res}, upper panel left, orange points), which has not
seen by any previous work on the bulge.

Regarding $\sigma_b$ (see Figure \ref{prev_res}, lower panel left), 
we found a large value that marginally follows the trend
of MS bulge stars, as seen by Kozlowski et al. (2006) in $\sigma_b$ vs. $l$.
Other authors' results agree as well, except for Rattenbury et al. 2007,
who observe systematic changes in $\sigma_b$, which are larger and offset from
the rest of the proper motion works.
M\'endez et al. (1996), (Figure \ref{prev_res} lower panel, black point),
who also studied proper motions in Plaut's Window, 
but using all stars with $V\leq 18$ 
and proper-motion errors less than 1 mas yr$^{-1}$,
found essentially the same result as ours.
This was initially surprising, since the M\'endez et al. (1996) sample
is certainly dominated by disk foreground stars, but their result
is just a convolution of the real disk velocity
dispersion along distance, plus the presence of real bulge stars,
and it is also affected by the proper motion errors of the
fainter stars. Indeed, a velocity dispersion of $\sim$ 30 km s$^{-1}$
for the thin disk foreground stars will show up with a proper motion dispersion of 
$\sim$ 3.2 mas yr$^{-1}$ and $\sim$ 1.6 mas yr$^{-1}$ for stars at 
around 2 and 4 kpc from the Sun, respectively. Thick disk
stars also contribute, although they are significantly less in number,
but they have two times larger dispersion than thin disk stars. 

In the case of the observed trends in Rattenbury et al. 2007, we would
like to call the attention to the fact that their selection of bulge stars,
an ellipse centered on the HB clump as seen in the $VI$ CMD, is 
a poor criterion since disk contamination in this part of the CMD
is rather important. Indeed the clump is produced by bulge stars, but
the presence of an important smooth distribution of disk giants in that same
magnitude and color range, undermines their selection and therefore their
the final results. The HB red clump is clearly visible in our data as well,
and we also studied the proper motion distribution of these stars. 
We found that their mean motion was in the mean point between the mean
motion of the bright blue nearby disk and the mean motion of the red distant bulge. 
This clearly reveals the strong mix of these populations
at this location in the CMD. We also obtained these mixed results
when selecting stars in the AGB red bump.
On top of that, except for the proper motion data obtained from space (see Table
\ref{pm_in_PW}), trustable ground-based proper motion requires an extended baseline
of at least a couple of decades, in order to achieve the accuracy needed
to properly assess the bulge kinematics. Numerous positions for a star
within a short period of time, as in the case of the OGLE-II data
used in Rattenbury et al. 2007, may improve its precision, but not its
accuracy. Therefore, we consider that their results might be affected
by systematics related to both, their selection criteria as well as 
their short baseline.

In the case of Kozlowski et al. (2006), the observed trends in dispersion for 
bulge MS stars, are explained as follows:
\begin{itemize}
\item[$\sigma_l$:]
Its value increases toward $b=0$ due to (i) Disk contamination or (ii) Higher rotation rate.
Any of these two factors can broaden $\sigma_l$.
\item[$\sigma_b$:]
Its value increases toward $l=0$ due to (iii) An increase in metal poor stars on the minor axis.
It has been observed that metal poor stars have larger $\sigma_b$
and $\sigma_l/\sigma_b$ closer to one (less anisotropy).
\end{itemize}
Of the three possible hypotheses, (i), (ii) and (iii) above, we will discuss 
disk contamination in more detail, since the other two topics - higher rotation rate
and the distribution of metal poor stars - are beyond the scope
of this paper. 

\subsection{Disk contamination in proper motions}

Disk contamination can be estimated from the distribution 
of observed proper motions, as long as all kinematic effects
involved are taken into account.
Previous studies did not deal with this issue, since they
only accounted for the proper motion shifts between bulge and foreground stars
as due to {\it disk kinematics}. In this paper we have done
our best effort to quantify all the observed trends in the 
proper motions, to better understand what they mean and where they
come from. So, for example, the effect of disk contamination 
on the measured bulge $\sigma_l$, when coming from nearby
stars can be easily guessed from our Figure \ref{pm_histo}:
disk star proper motions are biased towards positive values of $\mu_l$,
due to the reflex solar motion. If a bulge sample has nearby disk contamination, 
it will probably show a skewness in
its proper motion histogram and the measured dispersion $\sigma_l$ will be
artificially higher, like for example 
in Figure 6 of Kozlowski et al. (2006). Since our bulge sample
does not exhibit any skewness, this gives us confidence that
disk contamination in our sample is actually small.

This diagnosis is valid of course only for nearby disk stars, 
on which the reflex solar motion is the largest kinematic trend
affecting these stars. For more distant stars, beyond $\sim 2$ kpc,
we have to consider what Galactic rotation
does to their proper motion distribution, on both its mean value
as well as on its dispersion. To better understand this issue,
we must evaluate how differential galactic rotation projects onto
the proper motions along distance. For this, we selected the rotation curve
measured by Zabolotskikh, Rastorguev \& Dambis (2003),
who used a variety of young stellar systems plus HI and HII gas,
assuming $R_\odot=7.5$ kpc.
This value of $R_\odot$ is very close to $R_\odot=7.52$ kpc measured
by Nishiyama (2006) using HB red clump stars. 

The rotation curve in Figure 1 of Zabolotskikh, Rastorguev \& Dambis (2003),
can be easily transformed to observed proper motions $\mu_l$ vs. $V$,
by subtracting the Sun's rotational velocity, $V(R_\odot)=206\pm 10$ km s$^{-1}$,
and transforming those observed velocities into proper motions,
according to the distance. Extinction effects must also be taken into account.  
Figure \ref{rotcurve_pm} shows the reflex solar
motion (A03, long-dashed curve) and the galactic rotation (short-dashed curve) as seen in the observed proper motions,
individually and then added up (gray curve). The mean $\mu_l$ observed in our disk star sample, 
shifted by an amount to fit the gray curve, is also shown for comparison. 
Though it is clear that galactic rotation will shift the mean $\mu_l$ to
more negative values for stars fainter than $V\sim 16.7$, 
the final result once solar motion is considered,
can not reproduce the observed drop in the stellar proper motions
of these faint and therefore distant stars. 
One important issue here though, is that for galactocentric distances
$R<6$ kpc, Zabolotskikh, 
Rastorguev \& Dambis (2003) use only HI and HII data to measure the
rotation curve. It is well known that the gas disk in the Milky Way has
a smaller velocity dispersion and therefore a higher circular velocity than
the stellar disk. Similarly younger stellar populations have
more rotational support than older ones, which lag behind.
This is the so-called {\it asymmetric drift} 
in the solar neighborhood (see for example Vallenari et al. 2006), and has been also observed in external disk
galaxies (Vega Beltr\'an et al. 2001). Our sample of disk stars with $B-V\sim 0.7$ are mostly
dwarf G5 to K0-type, so it is expected that they have a 
higher velocity dispersion  and therefore a slower circular velocity
around the Galactic center, when compared to the gas disk.
This can explain why our disk stars exhibit a larger drop in their
observed mean $\mu_l$, than what is seen in the gas. 

If we were able to see deeper inside the 
Galaxy, a more negative mean $\mu_l$ would be observed,
down to the point where it could get close to the observed 
``negative'' mean  $\mu_l$ of the bulge 
(See Figures \ref{pm_histo} and \ref{pm_disk}). The proximity between the observed mean 
proper motion of the bulge and the disk, will depend of course on how
far we are observing from the Galactic rotation axis. 
It has also been observed that old disk stars increase
their velocity dispersion as they get closer to the Galactic center.
Lewis \& Freeman (1989) measured radial velocities for 600 old disk $K$ giants,
at different locations in $l$ for $2^o<|b|<5^o$ (See their Table I). Based on these data,
they computed the velocity dispersion along the azimuthal direction $\sigma_\varphi$,
as a function of galactocentric distance $R$. 
For example, for $R=3$ kpc, $\sigma_\varphi\sim 65$ km s$^{-1}$
($R_\odot=8.5$ kpc was assumed in their paper). 
For $l=0$, $\sigma_\varphi=4.74\sigma_l d$, with $d$ being distance.
Hence, following the example for $R=3$ kpc, $d=8.5-3.0=5.5$ kpc and $\sigma_l$ is $\sim 2.5$ mas yr$^{-1}$.
This value of disk dispersion is still smaller than the bulge dispersion
but not small enough to stand out as a secondary peak in a histogram 
of proper motions, particularly when the disk mean $\mu_l$ is also getting
closer to the bulge's mean motion.
Therefore, when sampling deep into these parts of the Galaxy, disk contamination
might actually decrease rather than increase the observed bulge $\sigma_l$,
as measured by eq. (\ref{spanhauer}). 
This problem can be of special importance in very small deep fields of view, 
when it is more probable to have disk
contamination from distant stars than from nearby. 

In order to further explore this idea, we decided to run
the Besancon Model of the Galaxy (Robin et al. 2003, 2004) 
on a homogeneous grid of points covering the pointings
studied in the Kozlowski et al. (2006) investigation. 
For the model data, we used the same constraints of their paper:
30''$\times$30'' field of view, 0.5$\leq$ V-I $\leq$3.0 and  18 $\leq$ I $\leq$ 21.5.
We also assumed an extinction suitable to reproduce the $I$ vs. $V-I$ CMD published in 
their paper. We found that contamination from disk stars in the sample selected,
ranged from 10\% to 15\%. It was observed indeed that for the 
fields closer to the Galactic plane, and therefore more heavily extincted,
the sample selection criterion of 18 $\leq$ I $\leq$ 21.5 allows
some of the disk stars brighter than the bulge turn off point to
make it into the sample, but in all cases more than 80\% of the disk stars
were located at distances $d>5$ kpc.
Disk stars alone showed dispersions typically 2 to 5 times smaller than those of
bulge stars and their mean motion was close enough to the
mean motion of the bulge, to make the whole sample $\sigma_l$
as much as 5\% to 10\% smaller than the real bulge dispersion in most cases,
when measured following eq. (\ref{spanhauer}). 
When examining the $\mu_l$ histograms though, it was possible to see in a few fields
closer to the galactic plane, a slight skewness that was indicative of the presence of
more than one single population. In general, the effect of the contaminanion
from distant disk stars
was to ``move'' the bulge mean motion to a more positive value, 
artificially reducing the real bulge $\sigma_l$, when using eq. (\ref{spanhauer}) for the whole sample.
These decrements represent
2 to 3 times the dispersion errors quoted by Kozlowski et al. (2006),
which means that they could - marginally - be observed when compared
to the dispersion of a truly pure bulge sample.
More important, disk contamination never increased the bulge dispersion,
and we did not find evidence of it producing
systematic changes in $\sigma_l$ or $\sigma_b$ as a function of $l$ or $b$, either.

Therefore, for these small deep fields of view, 
with a data selection as the one explained above, we could actually expect 
a reduction - if observable at all - rather than an increase, 
in the observed bulge $\sigma_l$, due to distant disk contamination,
since the nearby disk stars - which would broaden $\sigma_l$ - 
are very few compared to their far more numerous distant counterparts.
It is necessary to carefully evaluate the proper motion histograms,
and check for possible overlapping distributions of stellar proper motions,
as long as a good number of data points allows trustable statistics.
On the other hand, nearby disk contamination is easy to detect and
correct for, by examining the proper motion histograms and looking for
outliers in the positive side of the $\mu_l$ histogram. Also, taking note
of the extinction effects is key in order to avoid disk contamination,
particularly when selecting samples by fixed cuts in magnitude. 
Hence, we think that if these two factors had been considered more carefully 
in Kozlowski et al. (2006), the gradient in $\sigma_l$ might actually dissapear.
In any case, more observations towards the Galactic
plane should shed light on this issue, to check not only for
a gradient in the bulge rotation, which has been proposed as another possible
cause of this trend, but also to discard systematic
effects or errors.

\subsection{Final remarks}

The selection of bulge stars for kinematic studies is usually done
through photometry cuts, aiming to select a particular population 
that is known to belong to this part of the galaxy.
What population is chosen to study the bulge depends, between other things, on
the quality and depth of the photometric data, shallower photometry
allows a reliable identification of only the brightest features of
the bulge stellar population in the CMD. But the increasing amount
of data coming from CCD observations and surveys, 
is opening another possibility, studying the bulge faint MS stars.
Extreme care must be taken in this case, 
because such a selection may include either nearby intrinsically faint
stars, which will broaden $\sigma_l$ over large fields of view, 
or may include many distant disk stars which could decrease $\sigma_l$,
specially when observing deep into the galactic center.
In this last case, if proper motions are available for a large enough
number of stars, those can be used to separate bulge from foreground 
stars, particularly faint disk stars,
as was done in Kuijken \& Rich (2002), where only stars with
$\mu_l$ well in the negative side were selected as bulge stars. Such a kinematic selection can be
done, being of course aware of the biases that such sampling
may introduce in the data.

When looking for the right place to measure the bulge proper motion dispersion,
we need to deal with at least three interplaying factors:
disk contamination, bulge stellar density and obviously interstellar extinction.
The place with the higher density of bulge stars, i.e. the galactic plane
$l=0$, happens to have also the highest proportion of disk contamination and
extinction. Getting away from the plane to avoid
disk stars and dust, without loosing that many
bulge stars, is then the way to go. The minor axis actually looks like the
ideal location for such observations, though surely this
subject can be quantitatively determined either thru analytical models or numerical
simulations of the bulge and disk. In any case though, extinction maps
are the main tool to search and find low extinction windows, but it is advisable
to focus searches on those places where disk contamination is minimal.

As we have concluded before, our apparent high values of dispersion can be easily
explained by the closeness of our bulge sample. The line of sight at the Plaut's Window, 
$(l,b)=(0,-8)$, crosses the bulge so that we sample the highest stellar density at the 
tangent point of an isodensity curve. This tangent point is obviously closer than $R_\odot$,
and the farther you go off the Galactic plane, the closer you sample mean distance will be. 
Assuming that all bulge proper motion dispersions can be straightforwardly compared,
based on the assumption that all samples lie at the same mean distance,
may be a source of mistakes, even worst, a source of systematic errors.
Interestingly, one of the ways to measure the orientation and shape of the Galactic bar
was precisely by looking at the systematic changes in the observed magnitude
of red clump stars along Galactic longitude (Stanek et al. 1997, Cabrera-Lavers et al. 2007).
But none of the systematic changes observed by
Kozlowski et al. (2006) and Rattenbury et al. (2007) can be explained by the bar
orientation, since the trend in $\sigma_b$ vs $l$ goes in the wrong
direction. It is known that the bar is closer towards positive $l$,
therefore, we should observe (if possible at all) a decreasing $\sigma_b$ 
towards $l=0$, due to the increasing distance.

It is also very critical to understand the effects of the solar motion
and galactic rotation on both the mean proper motion and its dispersion,
since disk contamination has different effects according to distance.
In our case, for example, disk stars are mostly
nearby foreground, so in order to measure the relative motion of
the disk with respect to the bulge, we need to correct the reflex solar motion,
which in turn requires distances to be computed. Only for the MS disk
stars we may use photometric distances, which also happen to be not precise enough
to do so. Nevertheless, a first estimate of this can be obtained from the
proper motion histograms, as depicted in Figure \ref{pm_histo},
where the bulge sample stands out from the disk samples,
in the $\mu_l\cos{b}$ distribution.
For redder nearby disk stars though, no option is left to estimate distances.
If an accurate solar motion correction were feasible, it would be
possible to convert our proper motions, from relative to absolute,
for those disk stars having photometric distances.
Once we have completed our CCD photometry
we expect to properly examine this issue.
Last but not least, the presence of the long thin bar and its signature in the proper motion
space, is a wholly unexplored subject of study in its own right.
Therefore, we consider that more proper motion studies using RGB bulge stars,
which are easier to observe and photometrically to select, 
should be done at different locations around the bulge, before stating
any further conclusions regarding kinematical trends in the bulge.

\section{Conclusions}

We have measured proper motions of $\sim$ 21,000 stars brighter than $V\sim 20.5$
in the low extinction Plaut's Window, that covers 20' $\times$ 20'.
The catalog contains $\alpha, \delta$, l,b, relative proper motions
in Galactic coordinates with their respective errors, optical
photometry $V$, $B-V$ based on our photographic plates, and 2MASS ID's with their corresponding
near infrared magnitudes $J,H,K$ and errors, when available.
 
Using infrared 2MASS photometry we have separated disk and
bulge stars. Nearby disk dwarf stars clearly show the signature 
of the reflex solar motion. Our determinations of the 
bulge proper motion dispersion, though the highest measured in the bulge
so far, it is still in good agreement with previous results.
Such a high value is explained by the relatively close
bulge sample that we selected, i.e. the brightest RGB stars.
The mean distance to this sample has been estimated to be $6.37$ kpc, based on the
observed $K$ magnitude of the bulge HB red clump.
Based on this, the measured velocity dispersion is in good agreement with
the theoretical expected values for a steady-state triaxial bulge.
The observed values of proper motion anisotropy and metallicity distribution are 
limilar to previously published values.
No changes in dispersion were seen as function of metallicity.
An independent small sample of presumably mostly BHB bulge stars was also studied.
This sample showed a higher dispersion in both $l$ and $b$,
but the same anisotropy $\sigma_l/\sigma_b$ than the RGB bulge sample.
Our paper provides the first kinematic link ever between BHB and metal
rich bulge stars.

We also explored the effect that solar motion and galactic rotation have
in the observed mean proper motion and the proper motion dispersion,
particularly along galactic longitude, according to distance. We concluded that large field-of-view
shallow surveys of the bulge, mostly contaminated by nearby faint disk stars,
may observe an artificially high $\sigma_l$, due to the reflex solar motion component
in the observed stellar proper motions. On the other side, deeper small field-of-view
shots of the bulge, may suffer the opposite effect: a lower $\sigma_l$, due
to (i) small dispersion of disk stars and (ii) mean motion of disk stars gets closer
to the mean motion of bulge stars, due to galactic rotation effects.
Additionally, the fact that the Galactic bulge has a non negligible depth and orientation,
makes it important to consider distance effects, when comparing proper motion
dispersions along different locations in the bulge.

A catalog containing the relative proper motions will soon be available
at the Yale Astrometry Group web page at {\it http://www.astro.yale.edu/astrom/},
and also in the electronic edition of AJ. Samples of it can be seen in Tables 
\ref{plauts_catalog1} and \ref{plauts_catalog2}. 
We are now processing CCD observations of Plaut's Window, 
to obtain more accurate $UBVRI$ photometry, that will allow us to make full use of the
$\sim$21,000 precise proper motions.
This will then be the most complete, extended and precise
investigation of the Galactic bulge proper motions using ground based data, and its final
results will be close in precision to those using HST
data, not going as deep but covering a much wider area on the sky,
which allows us to simultaneosly address the study of several components
of the Milky Way.

\acknowledgments
This work has been supported by NSF grants AST-0407292, AST-0407293 and AST 0406884.
Ren\'e M\'endez acknowledges support by the Chilean Centro de Astrof\'{\i}sica
FONDAP (no. 15010003). 
We would like to express our appreciation to the observers who took the photographic
plates and made this investigation possible: Sidney van der Bergh, William Herbst,
Jeremy Mould, Steven Majewski, Oscar Sa\'a and Mario Cesco.
K. Vieira also express her thanks to Dr. Donald Terndrup who kindly gave
her the positional information on a set of M stars observed by Frogel et al. (1990).
This publication makes use of data products from the Two Micron All Sky Survey, which 
is a joint project of the University of Massachusetts and the Infrared Processing and 
Analysis Center/California Institute of Technology, funded by the National Aeronautics 
and Space Administration and the National Science Foundation.

\begin{deluxetable}{llccrcr}
\tablewidth{0pt}
\tablecaption{Photographic plate material\label{plates}}
\tablehead{
\colhead{Telescope}&
\colhead{Epoch}&
\colhead{Observer}&
\colhead{Emulsion}&
\colhead{\# of Plates}&
\colhead{Scale}&
\colhead{Centering Precision}\\
&&\colhead{\tablenotemark{a}}&&&\colhead{\tablenotemark{b}}&\colhead{\tablenotemark{c}}}
\startdata
KPNO   & 1972     & vdB      &  103aO,D       &  13 \hspace{0.5cm} & 12.70 &  18 \hspace{1.3cm} \\ 
Hale   & 1973     & vdB\&H   &  103aO,D       &   7 \hspace{0.5cm} & 11.12 &  86 \hspace{1.3cm} \\ 
duPont & 1979     & JRM      &  IIaO, IIIaF   &   5 \hspace{0.5cm} & 10.92 &  28 \hspace{1.3cm} \\ 
duPont & 1992-93  & SRM      &  IIaD, 103aO,D &  12 \hspace{0.5cm} & 10.92 &  28 \hspace{1.3cm} \\ 
CTIO   & 1993     & OSAA     &  IIIaJ         &   4 \hspace{0.5cm} & 18.59 &  19 \hspace{1.3cm} \\ 
YSO    & 1993     & MRC      &  103aO         &   2 \hspace{0.5cm} & 55.10 & 112 \hspace{1.3cm} \\ 
\enddata
\tablenotetext{a\;}{Observer: vdB$=$S. van der Bergh, vdB\&H$=$S. van der Bergh \&
W. Herbst, JRM$=$J. Mould, SRM$=$S. Majewski, OSAA$=$O. Sa\'a, MRC$=$M. Cesco.}
\tablenotetext{b\;}{Measured in arcsec mm$^{-1}$}
\tablenotetext{c\;}{Measured in milli-arcseconds}
\end{deluxetable}

\begin{deluxetable}{ccccccccc} 
\rotate
\tabletypesize{\small}
\tablewidth{0pt}
\tablecaption{Dispersion Measurements in the bulge\label{pm_in_PW}}
\tablehead{
\colhead{Author}&\colhead{l,b}&\colhead{Detector}&\colhead{FOV}&\colhead{Target}&
\colhead{\# of stars}&\colhead{$\Delta T$}&\colhead{Sample}&\colhead{Results} \\
\colhead{\tablenotemark{a}}&
\colhead{\tablenotemark{b}}&
\colhead{\tablenotemark{c}}&
\colhead{\tablenotemark{d}}&
\colhead{\tablenotemark{e}}&
\colhead{\tablenotemark{f}}&
\colhead{\tablenotemark{g}}&
\colhead{\tablenotemark{h}}&
\colhead{\tablenotemark{i}}}
\startdata
S92  & 1.02,-3.93 & plates & 2'5 - 6'       & NGC6522 & 427        & 33 & $B\leq19,\; (B-V)_0>1$ & 3.20,2.80 $\pm$ 1.00,1.00 \\
     &            &        & annulus        &         &            &    &                      & \\
M96  & 0.00,-8.00 & plates & 30'$\times$30' & PW      & $\sim$5000 & 21 & V$\leq$18   & 3.38,2.78 $\pm$ 0.03,0.03 \\
     &            &        &                &         &            &    &             & \\
Z01  & 5.25,-3.02 & WFPC2  & 2'$\times$2'   & NGC6553 & 1400       & 4  & V-I redder than  & 2.63,2.06 $\pm$ 0.29,0.21 \\
     &            &        &                &         &            &    & given line  & \\
     &            &        &                &         &            &    &             & \\
F02  & 1.14,-4.12 & WFPC2  & 2'$\times$2'   & NGC6528 & 500        & 6  & $V<19,\; V-I>1.6$ & 3.27,2.54 $\pm$ 0.27,0.17 \\
     &            &        &                &         &            &    &                     & \\
K02  & 1.13,-3.77 & WFPC2  & 2'$\times$2'   & BW      & 3252       & 6  & $\mu_l<-2,\; |\mu_b|<3$ & 2.94,2.63 $\pm$ 0.05,0.05\\
     &            &        &                &         &            &    &             & \\
K02  & 1.25,-2.65 & WFPC2  & 2'$\times$2'   & SGR I   & 3867       & 6  & $\mu_l<-2,\; |\mu_b|<3$ & 3.24,2.85 $\pm$ 0.05,0.05\\
     &            &        &                &         &            &    &             & \\
K06  & 2.10,-3.40 & ACS/HRC$+$ & 30"$\times$30" & BW  & 15254      & 4-8 & $18<I_{F814W}<21.5$ & 2.84,2.62 $\pm$ 0.11,0.04 \\
     &            & WFPC2  & $\times$35 fields  &     &            &     &                     &  \\
     &            &        &                &         &            &    &             & \\
R07  & 1.83,-2.92 & OGLE-II           & 14"$\times$57"     & $b=-4$     & 577736     & 4 & Red Clump Giants & 2.98,2.61 $\pm$ 0.03,0.03 \\
     &            & 2K$\times$2K CCD  & $\times$45 fields  & $-8<l<10$  &            &     &                     &  \\
     &            &        &                &         &            &    &             & \\
This  & 0.00,-8.00 & plates & 20'$\times$20' & PW     & 482        & 21  & Red Giant Branch        & 3.39,2.91 $\pm$ 0.11,0.09 \\
paper &            &        &                &        &            &     &            & \\
\enddata
\tablenotetext{a\;}{S92 $=$ Spaenhauer et al. 1992, M96 $=$ M\'endez et al. 1996, 
Z01 $=$ Zoccali et al. 2001, F02 $=$ Feltzing \& Johnson 2002, 
K02 $=$  Kuijken \& Rich 2002, K06 $=$ Kozlowski et al. 2006, R07 $=$ Rattenbury et al. 2007}
\tablenotetext{b\;}{Target position in Galactic coordinates. For K06 and R07, an average position of the 35 fields is shown.}
\tablenotetext{c\;}{WFPC2 $=$ HST Wide Field Planetary Camera 2. 
ACS/HRC$=$ HST Advanced Camera for Surveys/High Resolution Camera.}
\tablenotetext{d\;}{FOV$=$Field of View}
\tablenotetext{e\;}{PW=Plaut's Window, BW=Baade's Window}
\tablenotetext{g\;}{Epoch difference in years}
\tablenotetext{i\;}{Proper motion dispersion, $\sigma(\mu_l)\cos{b},\sigma(\mu_b)$ and their errors, all in mas$/$yr.
For K06 and R07, results are given for the mean position}
\end{deluxetable}

\clearpage

\begin{figure}
\begin{center}
\includegraphics[scale=0.8,angle=0,clip=true]{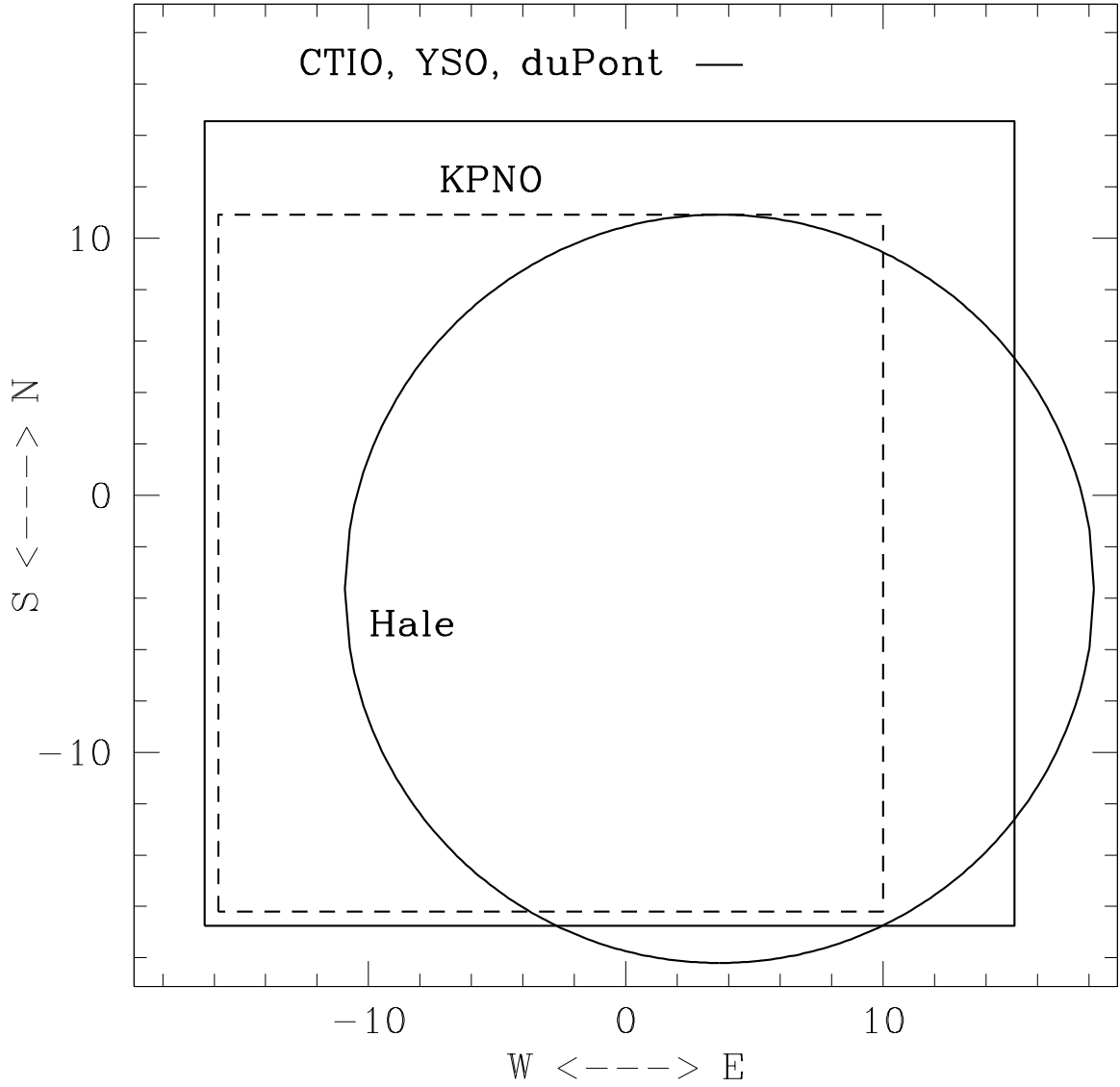}
\caption{Approximate spatial coverage and layout of the digitized plates.
Axes are measured in arcminutes.}
\label{fov_borders}
\end{center}
\end{figure}

\begin{figure}
\begin{center}
\includegraphics[scale=0.8,angle=0,clip=true]{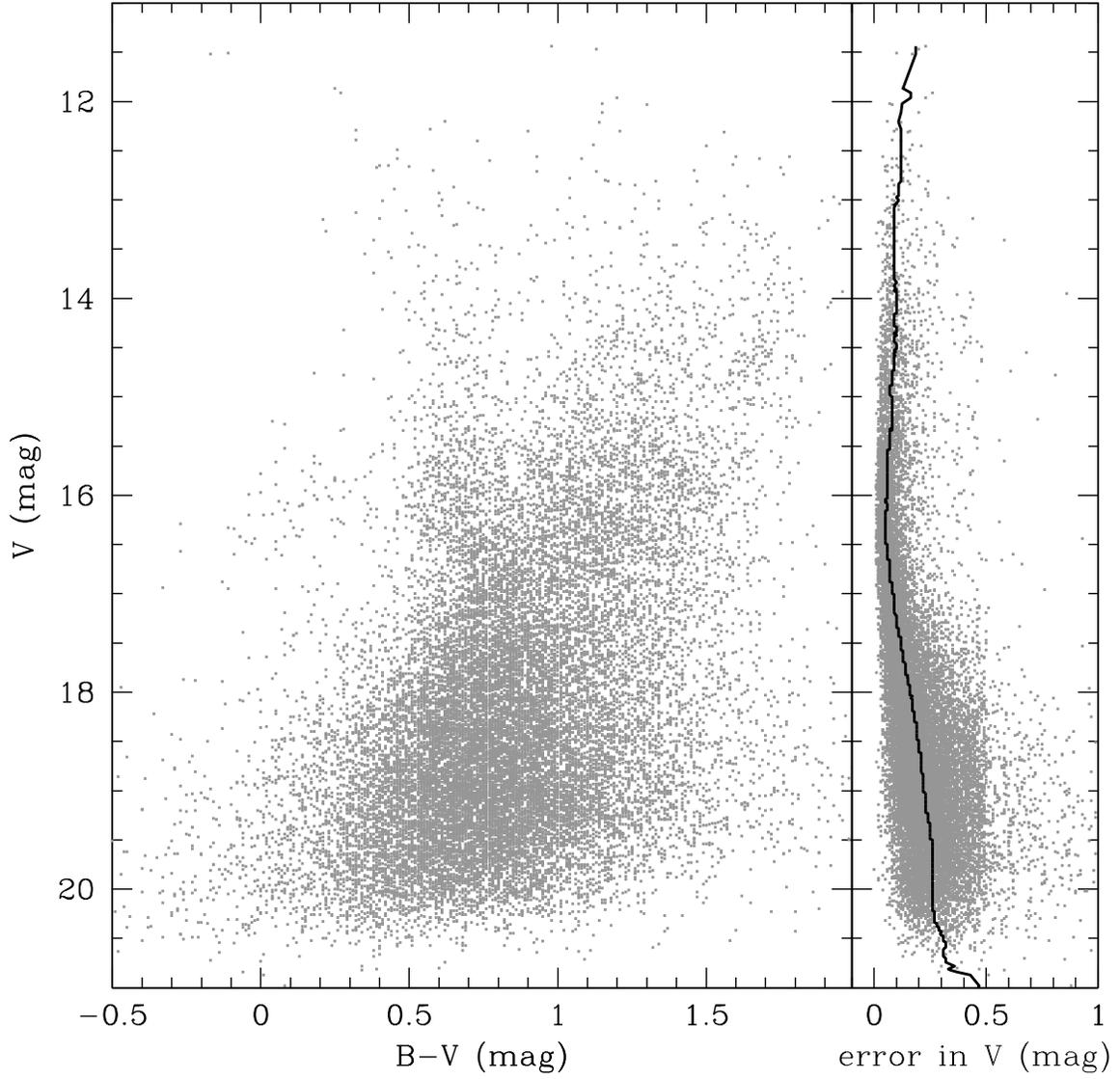}
\caption{Left: Optical CMD for the stars in Plaut's Window.
Right: Photometric error in $V$ vs. $V$ magnitude for the same stars.
The solid line is the median error computed
over a 0.4 mag interval.}
\label{cmd_err}
\end{center}
\end{figure}

\begin{figure}
\begin{center}
\includegraphics[scale=0.8,angle=0,clip=true]{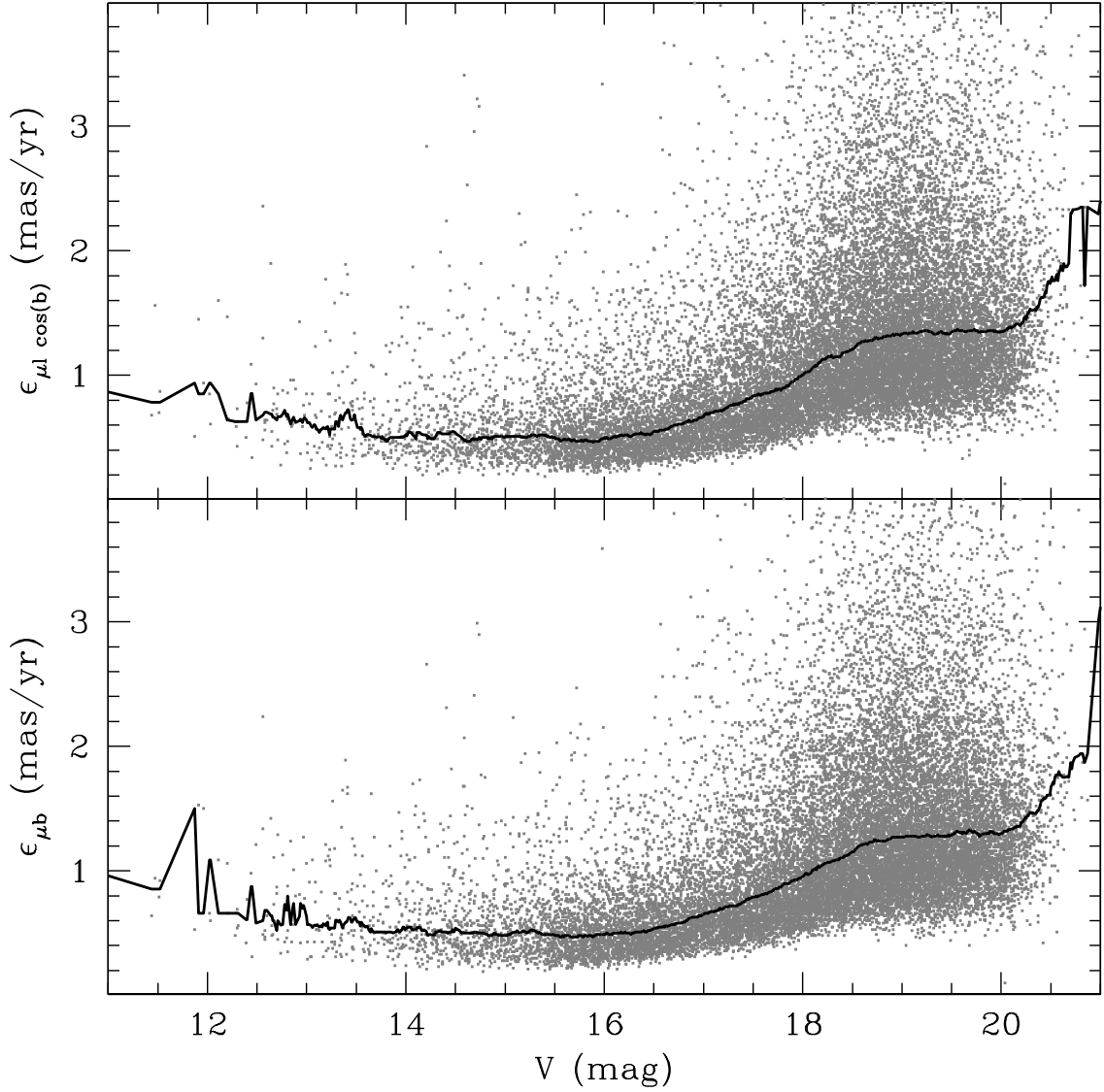}
\caption{Proper motion errors vs. $V$ magnitude, for stars in Fig. \ref{cmd_err}.
The solid line is the median error computed
over a 0.25 mag interval.}
\label{emum}
\end{center}
\end{figure}

\begin{figure}
\begin{center}
\includegraphics[scale=0.8,angle=0,clip=true]{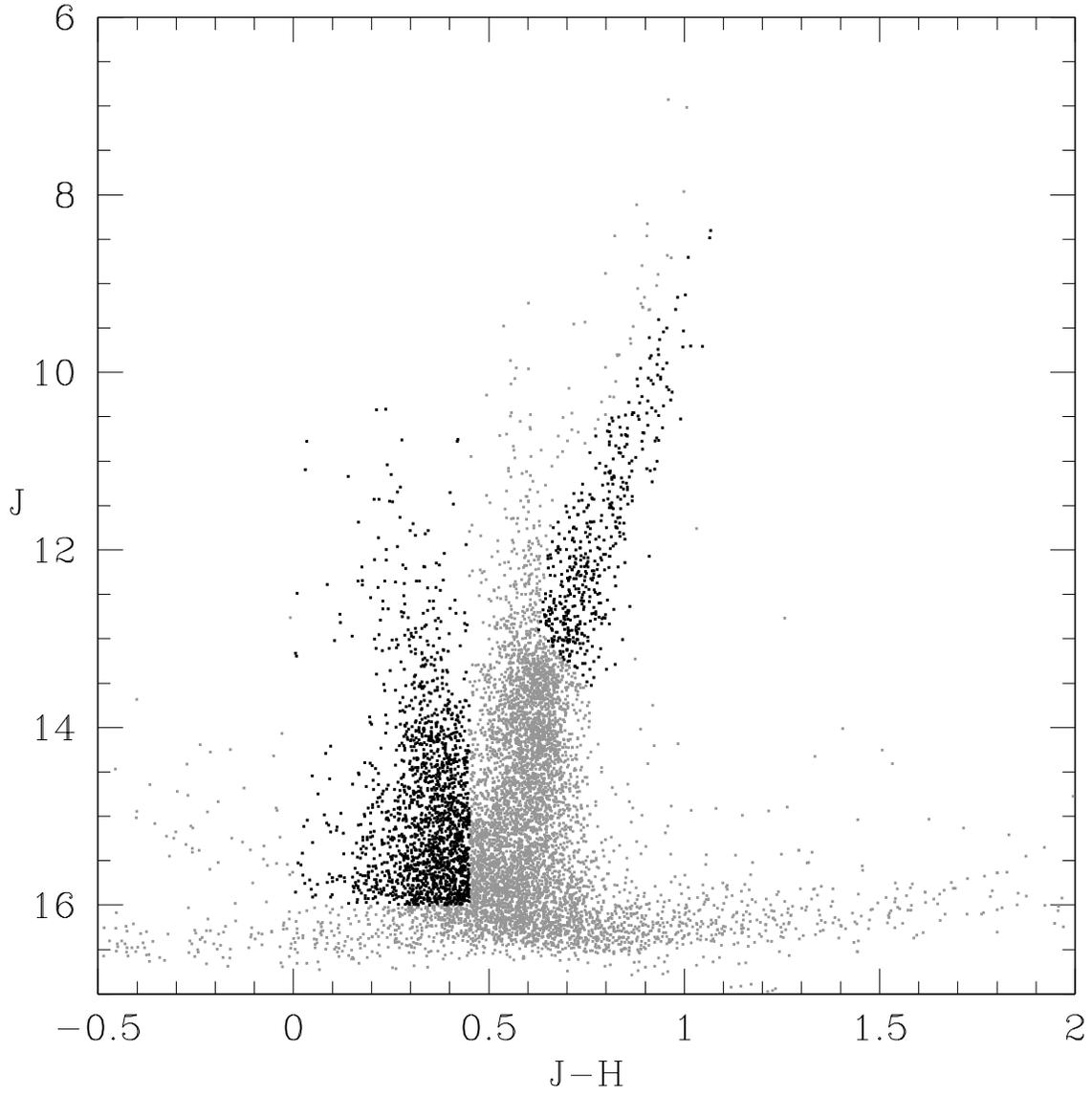}
\caption{2MASS Infrared CMD for stars in common with our study.
Disk stars were selected as those having $0.0\leq J-H\leq 0.45$ and $J\leq 16$.
Bulge stars are indicated by the black points on the RGB.}
\label{cmd_jh}
\end{center}
\end{figure}

\begin{figure}
\begin{center}
\includegraphics[scale=0.8,angle=0,clip=true]{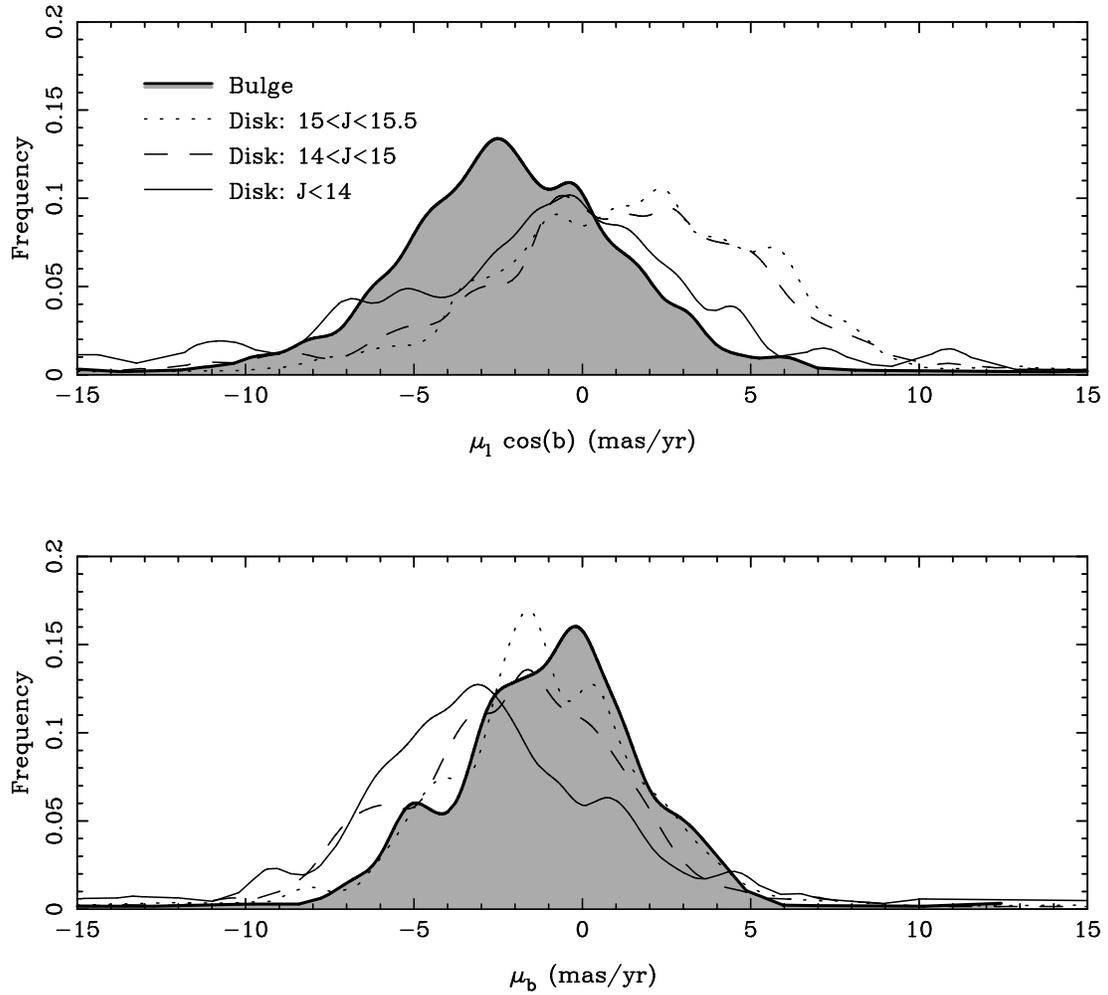}
\caption{Upper: Histogram of $\mu_l\cos{b}$ for three samples of disk
stars as well as the bulge stars. Lower: Same for $\mu_b$.
These are generalized histograms constructed using a gaussian kernel
of width 0.5 mas/yr.}
\label{pm_histo}
\end{center}
\end{figure}

\begin{figure}
\begin{center}
\includegraphics[scale=0.8,angle=0,clip=true]{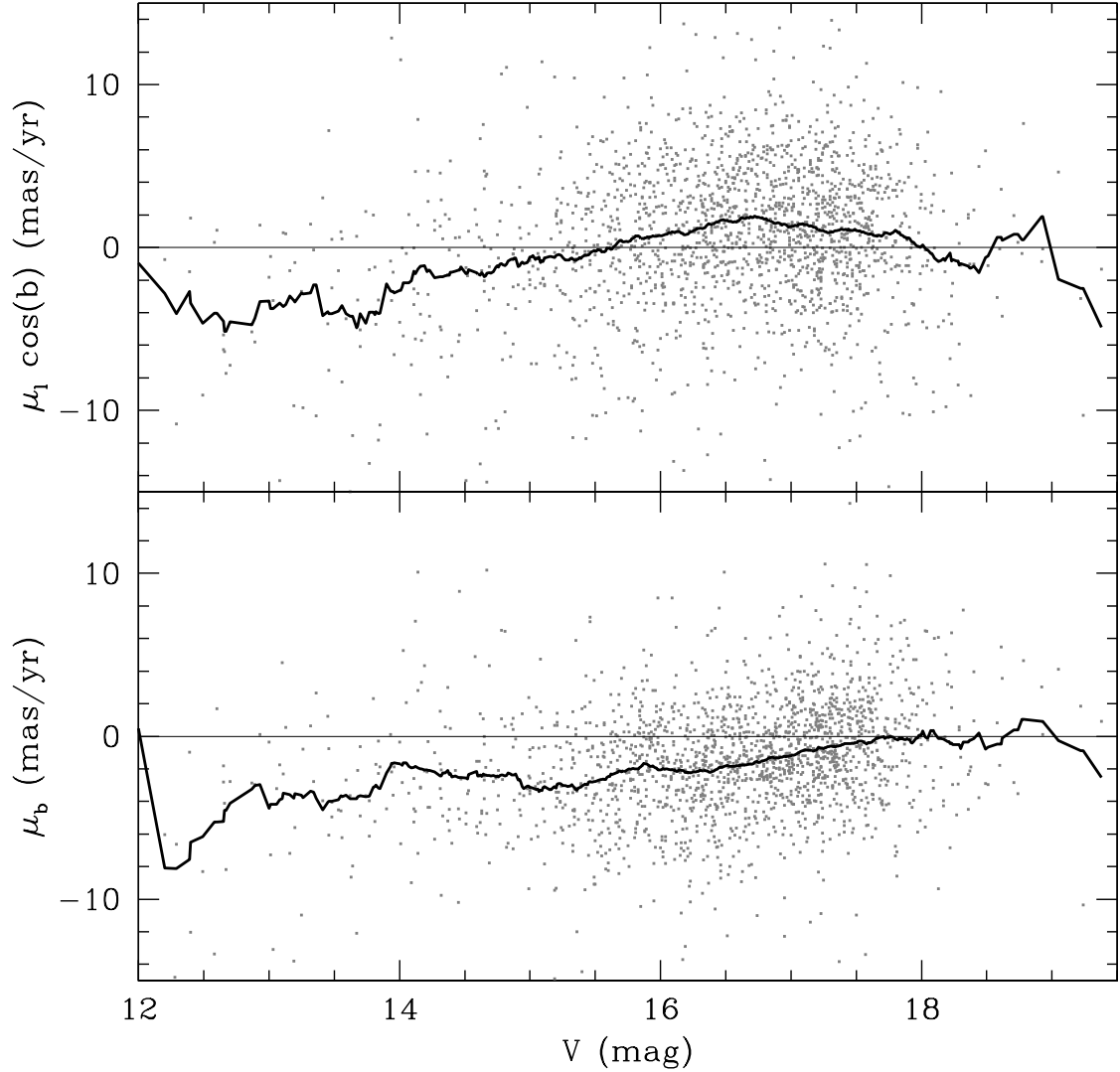}
\caption{Proper motion for the disk stars vs. $V$ magnitude (as
an indicator of distance). Solid line: Mean proper motion over
a 0.5 magnitude interval. The observed trend of larger proper motion
for the brighter (closer) stars, corresponds 
to the projection of the reflex solar motion.}
\label{pm_disk}
\end{center}
\end{figure}

\begin{figure}
\begin{center}
\includegraphics[scale=0.8,angle=0,clip=true]{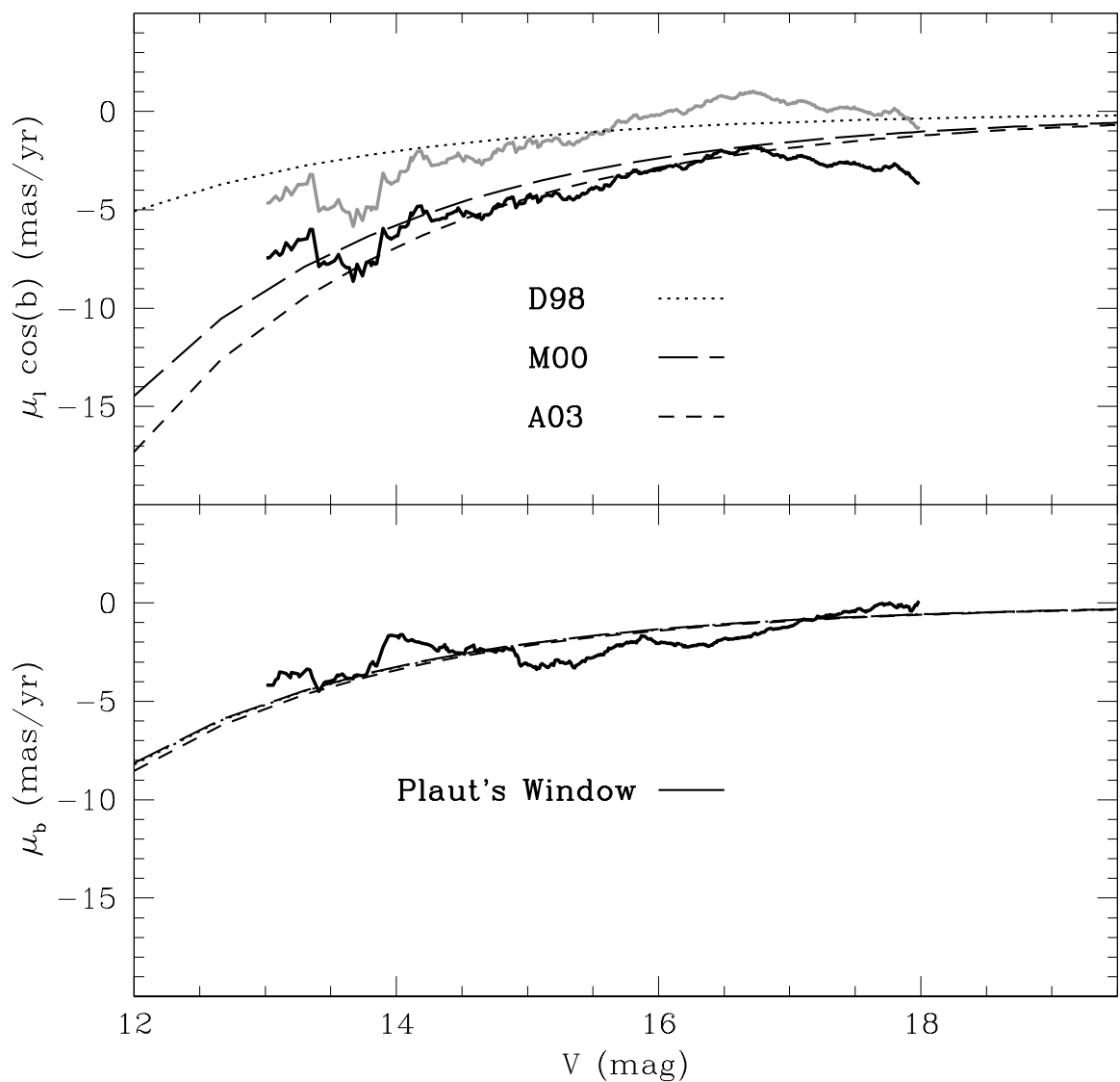}
\caption{Reflex solar motion projected on the observed mean proper motion.
Upper panel: The black and gray lines depict the solid 
line in the upper panel of Figure \ref{pm_disk}, 
plus an arbitrary shift in $\mu_l\cos{b}$ to fit Abad et al. (2003) and 
Dehnen \& Binney (1998) respectively. 
Lower panel: The black line depicts the solid line in the lower panel of Figure \ref{pm_disk}.
No shift was required in $\mu_b$ to fit the curves.} 
\label{solar_fit}
\end{center}
\end{figure}

\begin{figure}
\begin{center}
\includegraphics[scale=0.8,angle=0,clip=true]{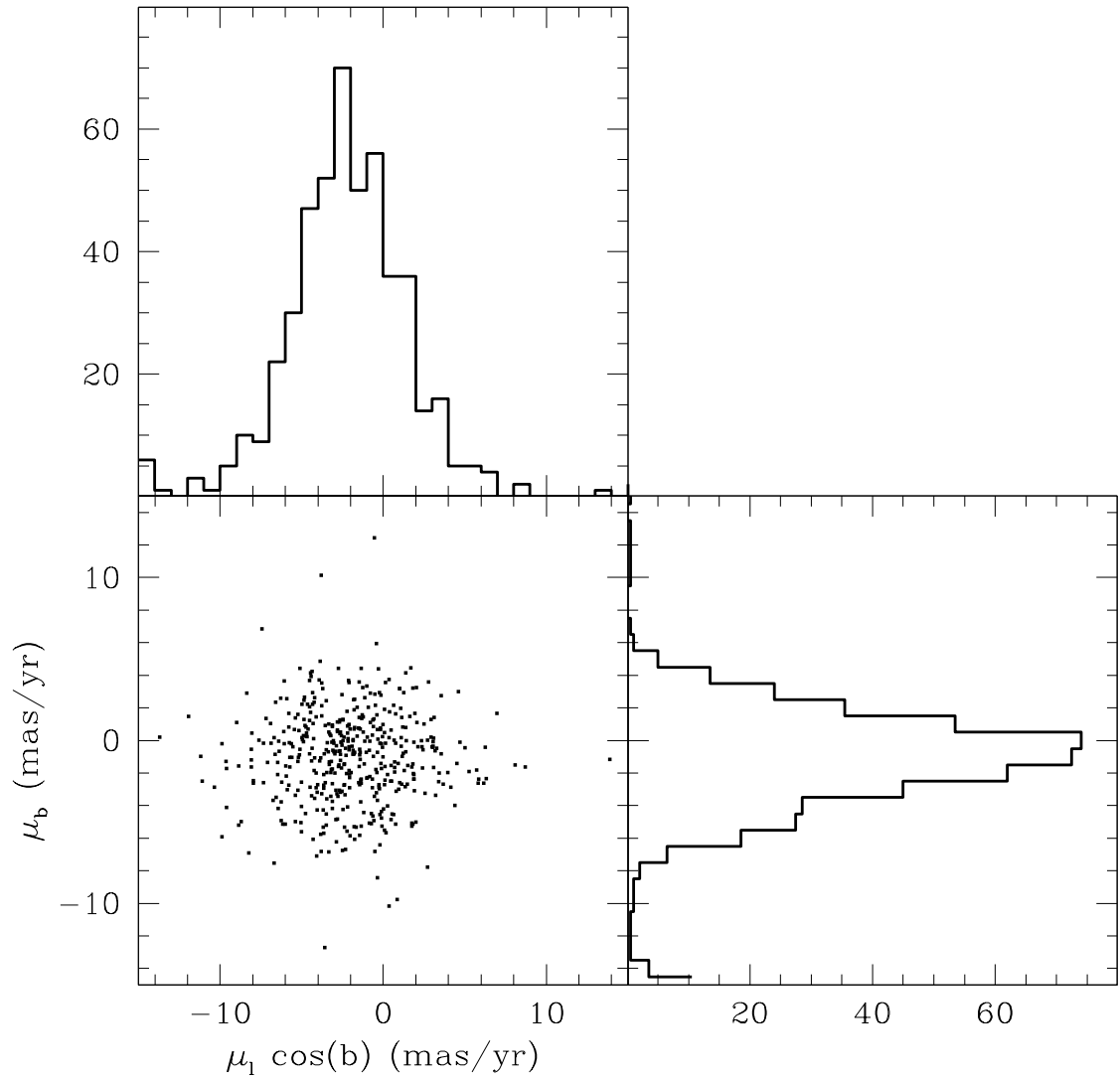}
\caption{Lower left: Vector-point diagram for the bulge stars.
Upper left: Histogram of $\mu_l\cos{b}$ for bulge stars.
Lower right: Histogram of $\mu_b$ for bulge stars.}
\label{pm_bulge}
\end{center}
\end{figure}

\begin{figure}
\begin{center}
\includegraphics[scale=0.8,angle=0,clip=true]{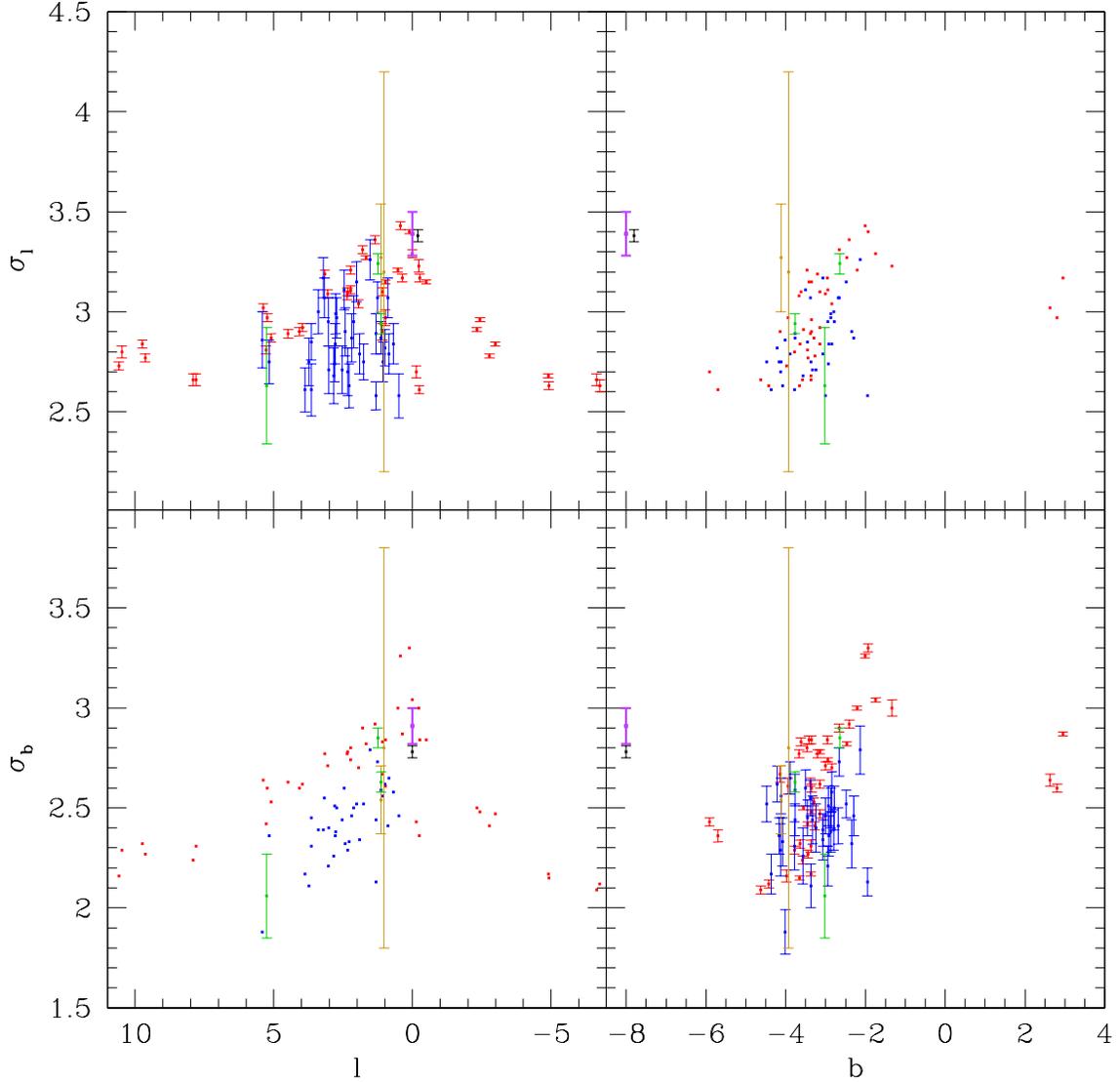}
\caption{Comparison of different proper motion investigations of the bulge. 
Results using MS bulge stars are in blue (Kozlowski et al. 2006)
and green (Zoccali et al. 2001, and Kuijken \& Rich 2002). Results
using RGB bulge stars are plotted in yellow (Spaenhauer et al. (1992), and Feltzing \& Johnson (2002)).
Results using the HB red clump (Rattenbury et al. 2007) are plotted in red.
The result from our study is plotted in bold purple. The M\'endez et al. (1996) result is plotted in black and
has been shifted slightly to the right in the upper panels, to distinguish it from our result.
Some error bars in the upper right and lower left panels are omitted for clarity purposes.}
\label{prev_res}
\end{center}
\end{figure}

\begin{figure}
\begin{center}
\includegraphics[scale=0.8,angle=0,clip=true]{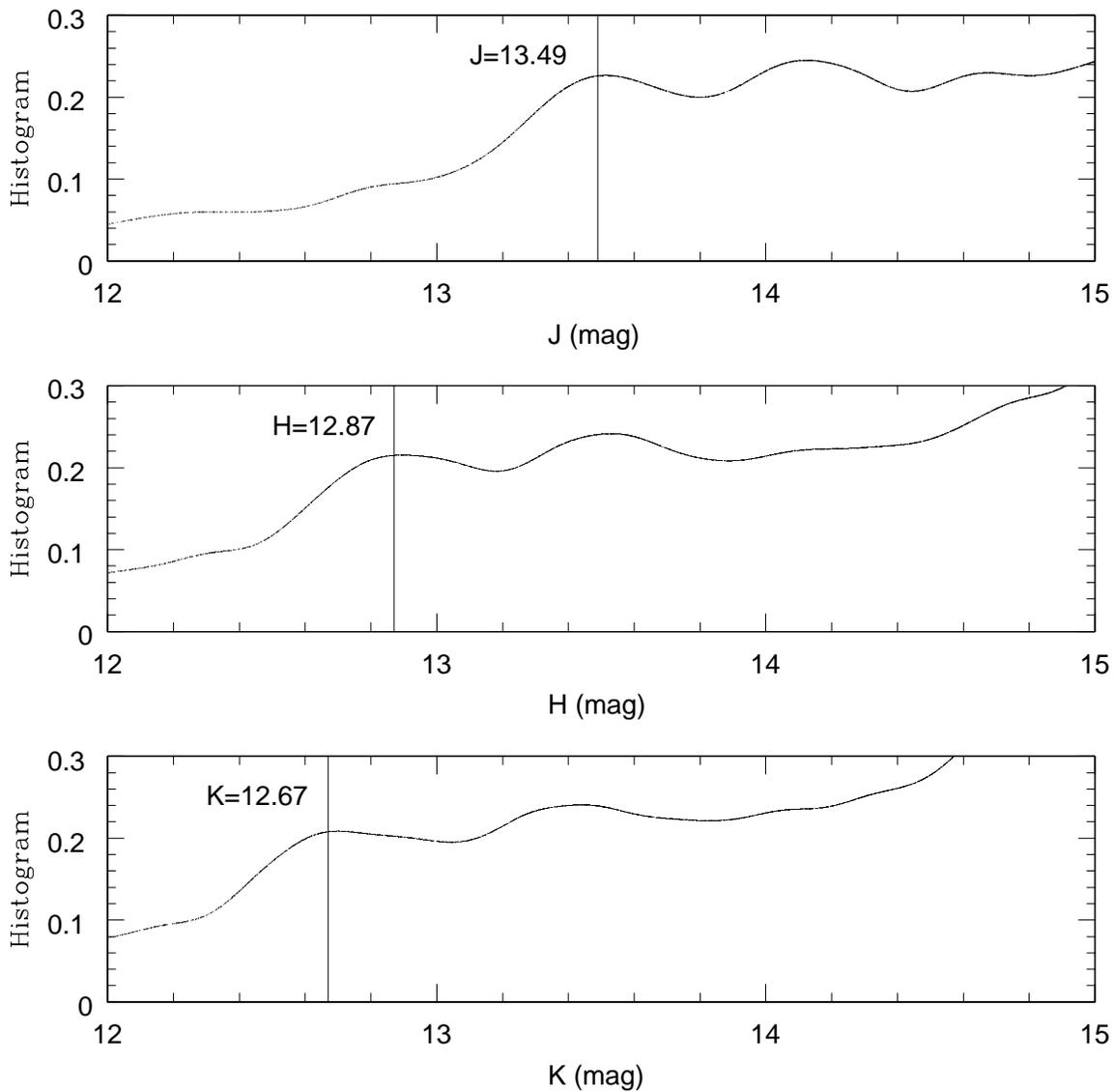}
\caption{Histogram of the Plaut's Window catalog in the $JHK$ 2MASS bands, 
for stars with $0.45 <J-H<0.8$. The HB Red clump is easily observed
as the brighter peak in the histograms.  
The vertical line marks the mean observed magnitude of the red clump stars.
The second fainter peak corresponds to the RGB bump. A gaussian kernel of
0.2 mags width was used to compute the histograms.}
\label{red_clumps}
\end{center}
\end{figure}

\begin{figure}
\begin{center}
\includegraphics[scale=0.8,angle=0,clip=true]{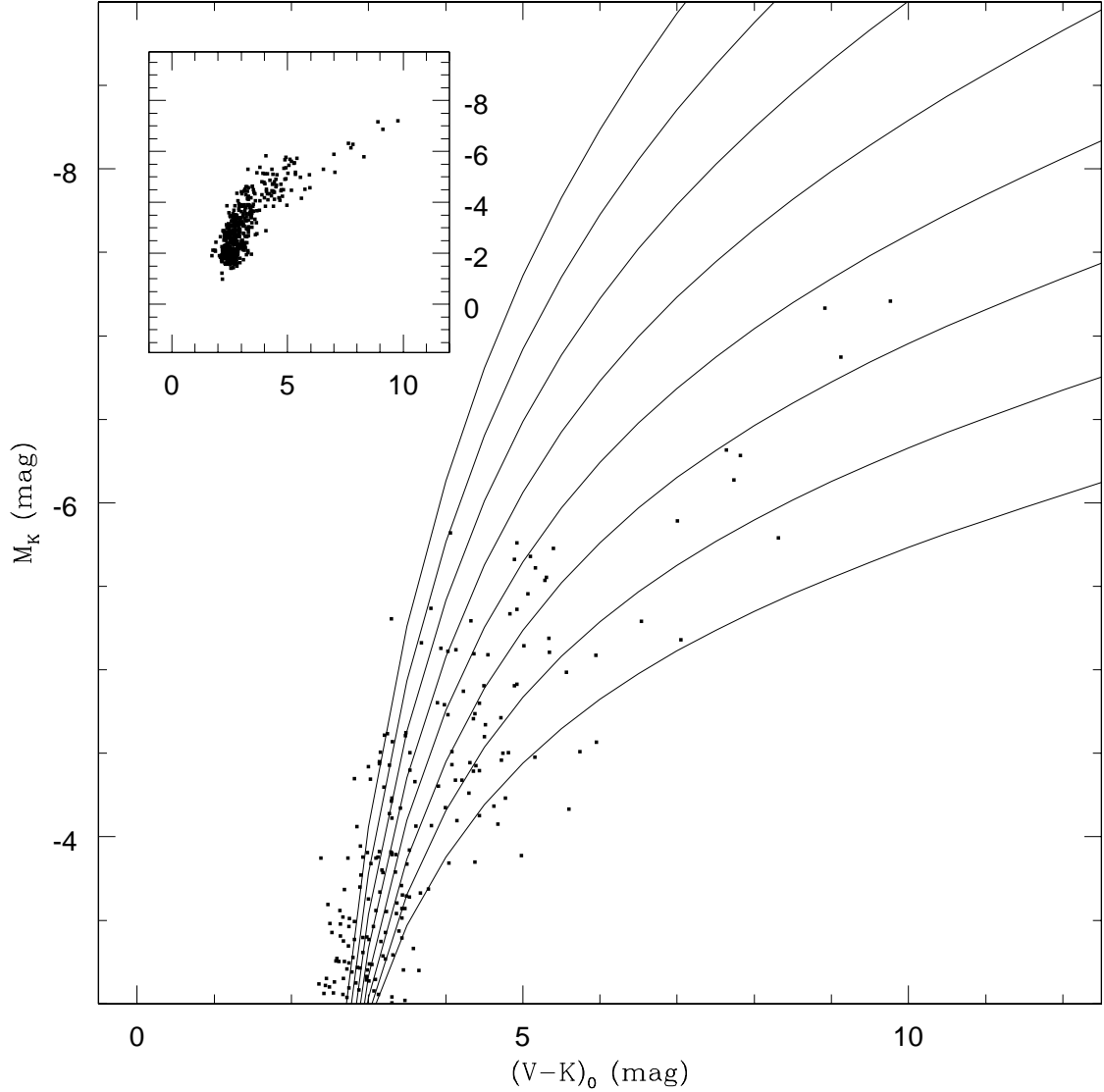}
\caption{$VK$ CMD of bulge giants compared with the analytical RGB templates for different
metallicities, from Zoccali et al. (2003). The leftmost curve corresponds to $[M/H]=-1.0$.
$[M/H]$ increases by 0.2 dex for the curves on the right. Insert on the upper left side is
a zoom-out of the same plot, showing all 482 bulge stars data. Only stars with $M_K>-4.5$ and
$(V-K)_0>2.8$ were used to determine the metallicity distribution, as shown in Figure \ref{metal}.}
\label{hyperbolas}
\end{center}
\end{figure}

\begin{figure}
\begin{center}
\includegraphics[scale=0.8,angle=0,clip=true]{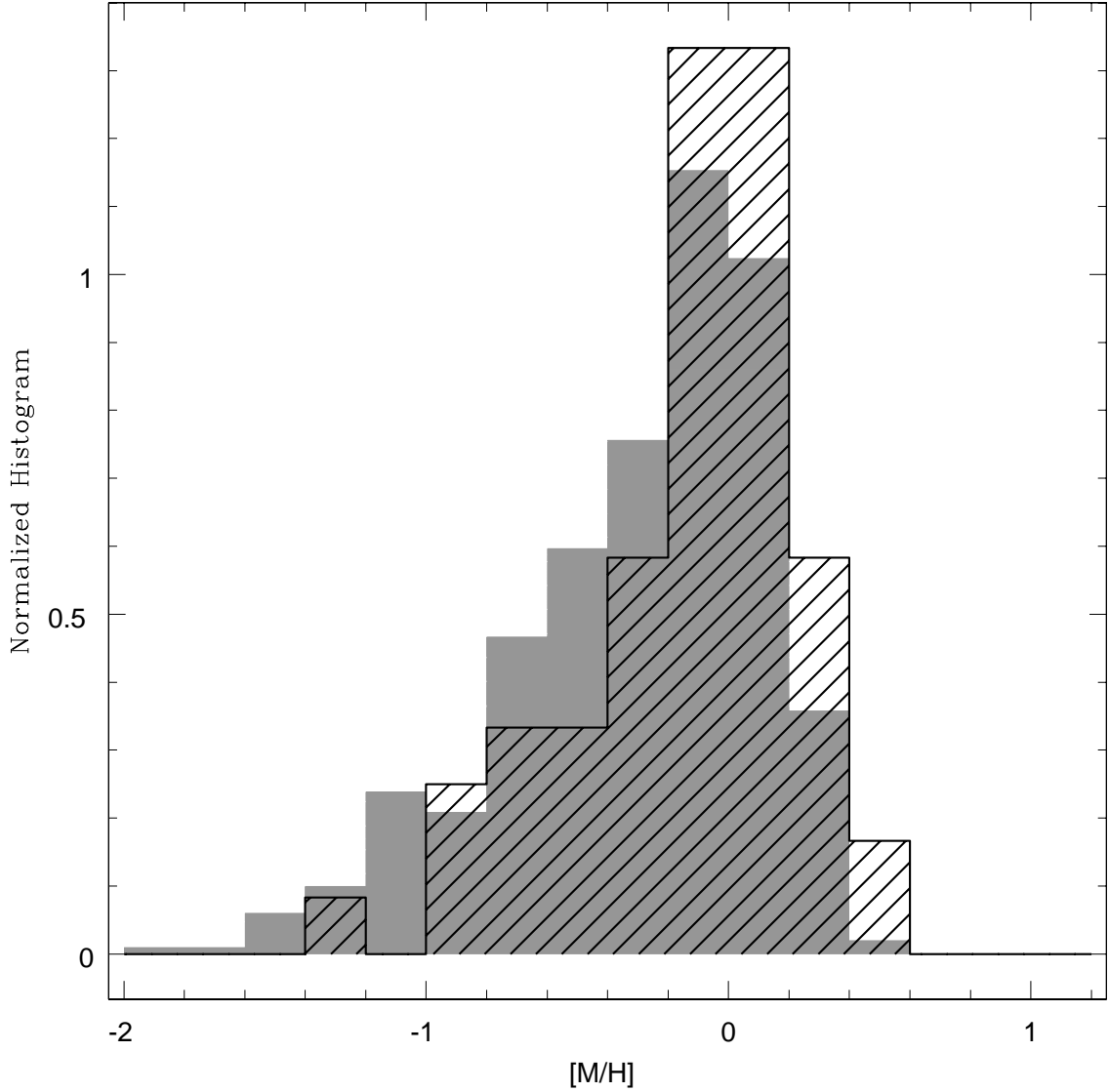}
\caption{Metallicity distribution of a subsample of 60 RGB bulge stars in the
Plaut's Window, as measured from Zoccali et al. (2003) hyperbolas calibration
from Figure \ref{hyperbolas}. Only stars with $M_K>-4.5$ and
$(V-K)_0>2.8$ were used to determine the metallicity distribution. In gray,
metallicity distribution of bulge giants from Zoccali et al. (2003).
Both histograms have been normalized for comparison.}
\label{metal}
\end{center}
\end{figure}

\begin{figure}
\begin{center}
\includegraphics[scale=0.8,angle=0,clip=true]{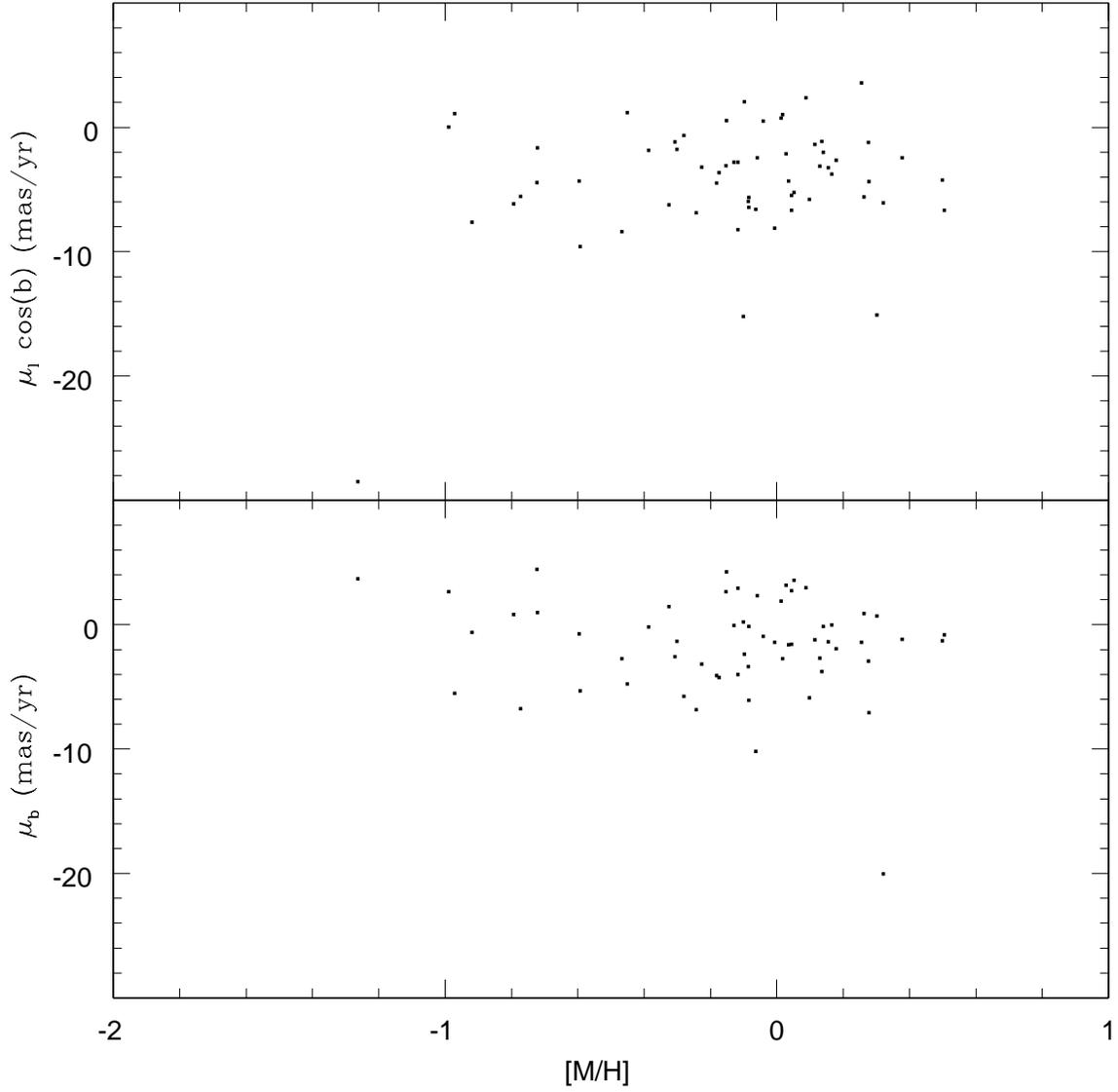}
\caption{Proper motion as a function of metallicity, for a subsample
of 60 RGB bulge stars in the Plaut's Window. No visible trend is noticed
in the proper motions, nor a significant change in dispersion between
metal-poor and metal-rich stars.}
\label{mu_mh}
\end{center}
\end{figure}

\begin{figure}
\begin{center}
\includegraphics[scale=0.8,angle=0,clip=true]{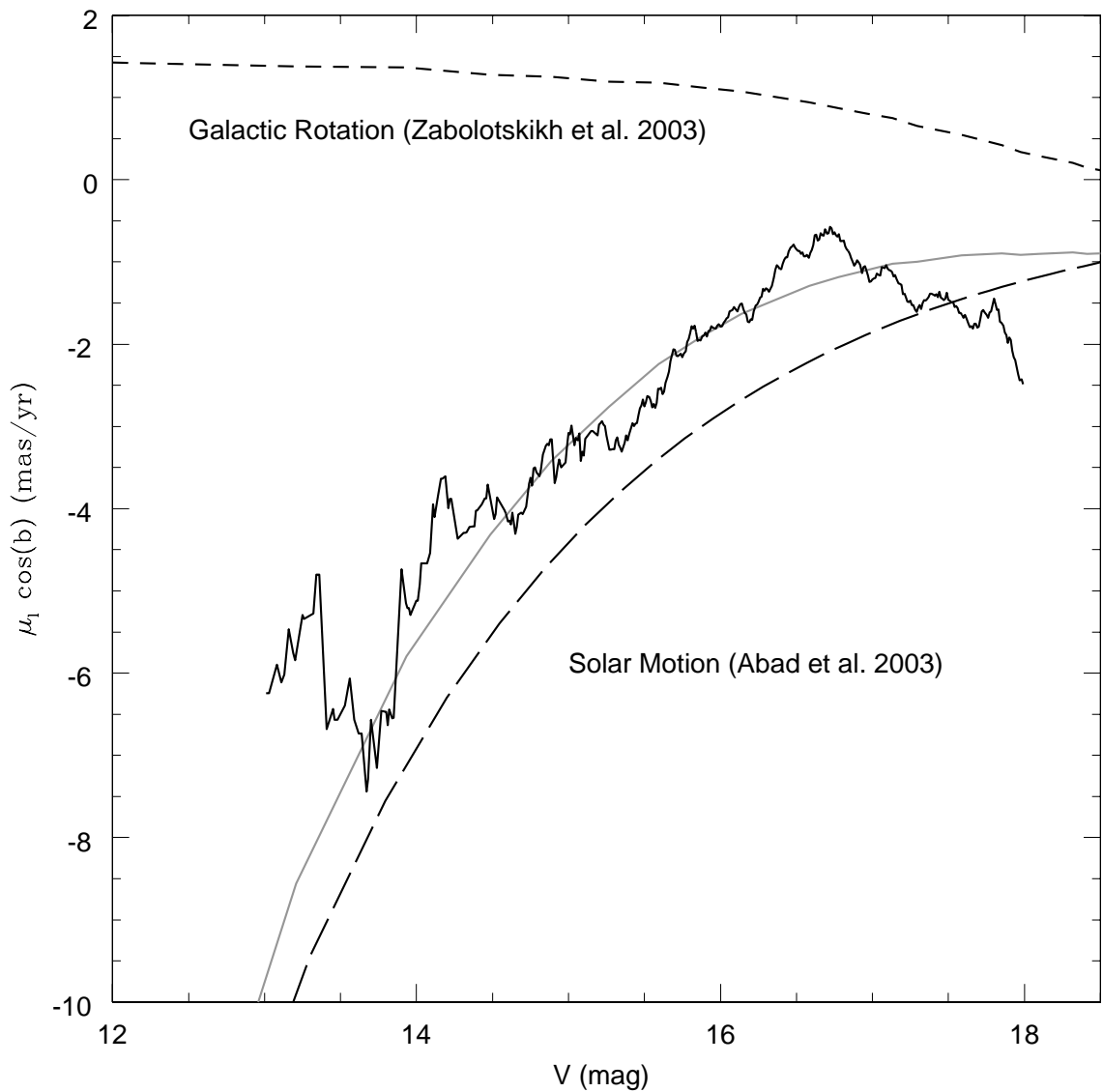}
\caption{Disk kinematics in the observed proper motion of our disk sample.
The long-dashed line is the reflex solar motion as measured by A03. 
The short-dashed line is the differential galactic rotation as observed
in $\mu_l$ as a function of $V$, based on Zabolotskikh, Rastorguev \& Dambis (2003)
rotation curve of the Milky Way. The gray line is the sum of the long- and the short-dashed lines.
The solid line is the mean $\mu_l$ observed over 0.5 mag intervals for our disk sample, 
shifted vertically to fit the gray line.}
\label{rotcurve_pm}
\end{center}
\end{figure}

\begin{deluxetable}{cccccrrrrr}
\rotate
\tablewidth{0pt}
\tablecaption{Plaut's Window Catalog of Proper Motions (Sample)\label{plauts_catalog1}}
\tablehead{
\colhead{ID}&
\colhead{RA}&
\colhead{DEC}&
\colhead{V}&
\colhead{B-V}&
\colhead{NP}&
\colhead{$\mu_l$}&
\colhead{$\mu_b$}&
\colhead{$\epsilon_{\mu_l}$}&
\colhead{$\epsilon_{\mu_b}$}\\
\colhead{number}&
\colhead{(deg)}&
\colhead{(deg)}&
\colhead{(mag)}&
\colhead{(mag)}&
&
\colhead{(mas yr$^{-1}$)}&
\colhead{(mas yr$^{-1}$)}&
\colhead{(mas yr$^{-1}$)}&
\colhead{(mas yr$^{-1}$)}\\
&
\colhead{\tablenotemark{a}} &
\colhead{\tablenotemark{a}} &
&
&
\colhead{\tablenotemark{b}}&
 & & & }
\startdata
 1 & 274.350370 & -32.985112 & 18.71 & 1.17 &  9 & -0.10\hspace{0.6cm} & -5.23\hspace{0.6cm} & 2.08\hspace{0.6cm} & 2.57\hspace{0.6cm} \\
 2 & 274.350386 & -33.006779 & 17.73 & 0.81 & 29 &-10.00\hspace{0.6cm} &  2.20\hspace{0.6cm} & 0.90\hspace{0.6cm} & 1.26\hspace{0.6cm} \\
 3 & 274.350411 & -33.004285 & 18.73 & 0.58 & 14 &  9.77\hspace{0.6cm} &  5.34\hspace{0.6cm} & 2.17\hspace{0.6cm} & 1.38\hspace{0.6cm} \\
 4 & 274.350529 & -32.998229 & 17.91 & 0.95 & 30 & -0.49\hspace{0.6cm} & -0.82\hspace{0.6cm} & 0.66\hspace{0.6cm} & 0.83\hspace{0.6cm} \\
 5 & 274.350649 & -32.987043 & 15.49 & 1.33 & 35 &  0.82\hspace{0.6cm} & -2.37\hspace{0.6cm} & 0.28\hspace{0.6cm} & 0.24\hspace{0.6cm} \\
\enddata
\tablenotetext{a\;}{J2000 (RA,DEC) as obtained from a solution with UCAC2}
\tablenotetext{b\;}{NP $=$ Number of plates on which the star appears}
\end{deluxetable}

\begin{deluxetable}{ccrrrrrrcr}
\rotate
\tablewidth{0pt}
\tablecaption{Plaut's Window stars with 2MASS data (Sample)\label{plauts_catalog2}}
\tablehead{
\colhead{ID}&
\colhead{2MASS ID}&
\colhead{J}&
\colhead{$\epsilon_J$}&
\colhead{H}&
\colhead{$\epsilon_H$}&
\colhead{K}&
\colhead{$\epsilon_K$}&
\colhead{Population}&
\colhead{Duplicate}\\
\colhead{number}&
\colhead{number}&
\colhead{(mag)}&
\colhead{(mag)}&
\colhead{(mag)}&
\colhead{(mag)}&
\colhead{(mag)}&
\colhead{(mag)}&
\colhead{index}&
\colhead{match ID}\\
&&&&&&&& 
\colhead{\tablenotemark{a}}&
\colhead{\tablenotemark{b}}}
\startdata
  18104 & 18184517-3242053 & 15.906 & 0.065 & 15.471 & 0.100 & 15.501 & 0.000 & 2 &       \\
  18106 & 18184517-3245568 & 15.854 & 0.098 & 14.238 & 0.000 & 14.275 & 0.000 & 0 & 18116 \\
  18111 & 18184518-3300268 & 14.756 & 0.062 & 14.388 & 0.089 & 14.462 & 0.093 & 2 &       \\
  18116 & 18184517-3245568 & 15.854 & 0.098 & 14.238 & 0.000 & 14.275 & 0.000 & 0 & 18106 \\
  18121 & 18184521-3241494 & 11.684 & 0.022 & 10.860 & 0.022 & 10.627 & 0.019 & 1 &       \\
\enddata
\tablenotetext{a\;}{Population index: 0=Undetermined, 1=Bulge RGB, 2=Disk MS.
An index value of 0 means that no population was explicitly assigned to the star}  
\tablenotetext{b\;}{ID of another Plaut's star that was cross-referenced to the same
2MASS star. Our frames have better resolution than 2MASS.}
\end{deluxetable}


\begin{references}

\reference{} Abad, C. et al. 2003, A\&A, 397, 345

\reference{} Bertin, E. \& Arnouts, S. 1996, A\&AS, 117, 393

\reference{} Beaulieu, S. et al. 2000, AJ, 120, 855

\reference{} Blaauw, A. 2004, ARA\&A, 42, 1

\reference{} Blanco, V.~M. 1988, AJ, 95, 1400

\reference{} Blanco, V.~M. \& Terndrup, D.~M. 1989, AJ, 98, 843

\reference{} Cabrera-Lavers A. et al. 2007, astro-ph/0702109v1 (submitted to A\&A)

\reference{} Cardelli, J.~A., Clayton, G.~C. \& Mathis, J.~S. 1989, AJ, 345, 245

\reference{} Carpenter J.~M. 2001, AJ, 121, 2851

\reference{} Chiu, L. 1976, PASP, 88, 803

\reference{} Cudworth, K.~M. \& Rees, R.~F. 1991, PASP, 103, 400

\reference{} Cutri, R.~M. et al. 2003, 2MASS All Sky Catalog of point sources (Pasadena: IPAC/Caltech)

\reference{} Dehnen, W. \& Binney J. 1998, MNRAS, 298, 387

\reference{} Dominici, T.~P. et al. 1999, A\&AS, 136, 261

\reference{} Dwek E. et al. 1995, ApJ, 445, 716

\reference{} Feltzing, S. \& Johnson, R.~A. 2002, A\&A, 385, 67
 
\reference{} Ferraro, F.~R., Valenti, E. \& Origlia, L. 2006, AJ, 649, 243
 
\reference{} Frogel, J.~A. et al. 1990, AJ, 353, 494
 
\reference{} Ghez, A.~M. et al. 2005, ApJ, 620, 744
 
\reference{} Girard, T.~M. et al. 1989, AJ, 98, 227

\reference{} Girard, T.~M. et al. 1998, AJ, 115, 855
 
\reference{} Guo, X. et al. 1993, AJ, 105, 2182
 
\reference{} Izumiura, H. et al. 1994, ApJ, 437, 419

\reference{} Izumiura, H. et al. 1995, ApJ, 453, 837

\reference{} Kohoutek, L. \& Pauls, R. 1995, A\&AS, 111, 493

\reference{} Kozhurina-Platais, V. et al. 1995, AJ, 109, 672

\reference{} Kozlowski, S. et al. 2006, MNRAS, 370, 435

\reference{} Kuijken, K. \& Rich, R.~M. 2002, AJ, 124, 2054

\reference{} Lee, J.~F. \& van Altena W. 1983, AJ, 88, 1683

\reference{} Lewis, J.~R. \& Freeman, K.~C. 1989, AJ, 97, 139

\reference{} M\'endez, R.~A. et al. 1996, ASP Conf. Series, 102, 345

\reference{} Mignard, F. 2000, A\&A, 354, 522

\reference{} Minniti, D. 1996, ApJ, 459, 579

\reference{} Ng, Y.~K. \& Bertelli, G. 1996, A\&A, 315, 116

\reference{} Nishiyama, S. et al. 2006, astro-ph/0607408v1 (accepted to ApJ)  

\reference{} Oort, J. \& Plaut, L. 1975, A\&A, 41, 71

\reference{} Paczynski, B. \& Stanek, K.~Z. 1998, ApJ, 494, L219

\reference{} Perryman, M.~A.~C. et al. 1997, A\&A, 323, L49

\reference{} Plaut, L. 1973, A\&AS, 12, 351

\reference{} Rattenbury, N.~J. et al. 2007, arXiv:0704.1619 v1

\reference{} Rich, R.~M. 1990, ApJ, 362, 604

\reference{} Rich, R.~M. 1998, in IAU Symp. 184, The Central Regions of the Galaxy and Galaxies,
             ed. Y. Sofue, (Dordrecht: Kluwer), 11

\reference{} Rich, R.~M. et al. 2007, AJ, 658, L29

\reference{} Robin, A.~C. et al. 2003, A\&A, 409, 532

\reference{} Robin, A.~C. et al. 2004, A\&A, 416, 157

\reference{} Sadler, E.~M.,  Rich, R.~M. \& Terndrup, D.~M. 1996, AJ, 112, 171

\reference{} Soto, M., Rich, R.~M. \& Kuijken, K. 2006, astro-ph/0611433v1 (submitted to ApJ Letters)

\reference{} Tiede, G.~P. \& Terndrup, D.~M. 1997, AJ, 113, 321

\reference{} Tiede, G.~P. \& Terndrup, D.~M. 1999, AJ, 118, 895

\reference{} Tyson, N.~D. \& Rich, R.~M. 1991, ApJ, 367, 547

\reference{} Sparke, L. \& Gallagher, J. 2000, Galaxies in the Universe (Cambridge: 
             Cambridge University Press)

\reference{} Spaenhauer, A., Jones, B.~F. \& Whitford, A.~E. 1992, AJ, 103, 297

\reference{} Stanek, K. Z. et al., 1997, AJ, 477, 163

\reference{} Stanek, K. Z. \& Garnavich, P. M. 1998, ApJ, 503, L131

\reference{} Terndrup, D.~M. 1988, AJ, 96, 884

\reference{} Terndrup, D.~M., Sadler, E.~M. \&  Rich, R.~M. 1995, AJ, 110, 1174

\reference{} Udalski, A. et al. 1998, Acta Astronomica, 48, 1

\reference{} Vallenari, A. et al. 2006, A\&A, 451, 125

\reference{} van den Bergh, S. \& Herbst, E. 1974, AJ, 79, 603

\reference{} Vega Beltr\'an, J,~C. et al. 2001, A\&A 374, 394

\reference{} Zabolotskikh, M. V., Rastorguev, A. S. \& Dambis, A. K. 2003,
Communications of the Konkoly Observatory, Hungary. Proceedings of the conference: 
``The interaction of stars with their environment II'', held at the E\"{o}tv\"{o}s 
Lori\'and University, Budapest, Hungary, May 15-18, 2003, ed. Cs. Kiss, M. Kun, 
V. K\"{o}nyves, p. 167-172

\reference{} Zhao, H. 1996, MNRAS, 283, 149

\reference{} Zhao, H., Rich R.~M. \& Biello, J. 1996, AJ, 470, 506

\reference{} Zijlstra, A.~A., Acker, A. \& Walsh, J.~R. 1997, A\&AS, 125, 289

\reference{} Zoccali, M. et al., 2001, AJ, 121, 2638

\reference{} Zoccali, M. et al., 2003, A\&A, 399, 931

\reference{} Zoccali, M. et al., 2006, astro-ph/0609052v2 (accepted to A\&A)

\end{references}
\end{document}